\pdfoutput=1
\documentclass[10pt,twocolumn,pre, floatfix]{revtex4-1} 
\usepackage{times}
\usepackage{amsmath}
\usepackage{float}
\usepackage{bm}
\usepackage{gensymb}
\usepackage[dvipsnames]{xcolor}
\usepackage{graphicx}
\usepackage{amssymb}
\usepackage{url,hyperref}

\newcommand{\mlm}[1]{\textcolor{black}{#1}}
\newcommand{\tah}[1]{\textcolor{black}{#1}}
\newcommand{\cds}[1]{\textcolor{black}{#1}}

\begin{document}

\title{\mlm{Optimizing properties on the critical rigidity manifold of underconstrained central-force networks}}

\author{Tyler Hain}
\affiliation{Department of Physics and BioInspired Institute, Syracuse University, Syracuse NY 13210}

\author{Chris Santangelo}
\email[]{cdsantan@syr.edu}
\affiliation{Department of Physics and BioInspired Institute, Syracuse University, Syracuse NY 13210}

\author{M. Lisa Manning}
\email[]{mmanning@syr.edu}
\affiliation{Department of Physics and BioInspired Institute, Syracuse University, Syracuse NY 13210}

\begin{abstract}
\mlm{Our goal is to develop a design framework for multifunctional mechanical metamaterials that can tune their rigidity while optimizing other desired properties. Towards this goal, we first demonstrate that underconstrained central force networks possess a critical rigidity manifold of codimension one in the space of their physical constraints. We describe how the geometry of this manifold generates a natural parameterization in terms of the states of self-stress, and then use this parameterization to numerically generate disordered network structures that are on the critical rigidity manifold and also optimize various objective functions, such as maximizing the bulk stiffness under dilation, or minimizing length variance to find networks that can be self-assembled from equal-length parts. This framework can be used to design mechanical metamaterials that can tune their rigidity and also exhibit other desired properties.}

\end{abstract}

\maketitle

\section{Introduction}

Mechanical metamaterials \mlm{are designed with specific meso-scale structures that imbue them with macroscopic mechanical properties or behaviors distinct from those of the material} from which they are constructed \cite{goh_nanolatticed_2021, wei_multi-bionic_2022}. \cds{The design space for such materials is vast,} 
\mlm{yet designed materials cannot match the breadth and versatility of evolved biological materials, which} exhibit an incredible ability to adapt their mechanical properties in response to external stimulus. For example, confluent tissues in developing embryos undergo controlled fluidization to facilitate the large-scale shape changes associated with development \cite{mongera_fluid--solid_2018, bi_density-independent_2015, bi_motility-driven_2016}. \cds{On the other hand,} biopolymer networks of crosslinked fibers can be very soft for small amounts of strain, but suddenly increase their stiffness by orders of magnitude at a particular critical strain \cite{mackintosh_elasticity_1995, storm_nonlinear_2005, sharma_strain-controlled_2016, rens_micromechanical_2018}.
While many metamaterials are designed as regular lattices of modular unit cells, \cds{biological materials are often} disordered, \cds{and consequently} have the potential for a wider space of functionality as well as increased mechanical robustness in comparison to ordered structures, for example by avoiding shear banding and fracture \cite{zaiser_disordered_2023}.
\cds{Similar adaptability and functionality in a metamaterial could be achieved by designing structures poised near the rigidity transition.} \cds{Yet the question of how to} design disordered mechanical metamaterials that take advantage of rigidity transitions,
so that small changes to the geometry of the structure significantly change the macroscopic mechanical properties \mlm{like biological examples}, \cds{remains completely open.}

\cds{In this paper, we introduce a method to completely enumerate and explore the entire space of \emph{geometric} rigidity transitions of large networks.}
\mlm{Past work has shown that underconstrained tissue and biopolymers} can be driven to a critical configuration by applying sufficient strain~\cite{merkel_minimal-length_2019}, but it has not been clear exactly how common these critical configurations are or whether they exhibit any emergent order or symmetry, especially in large disordered systems. In addition, the critical configuration that results from straining to the critical point is an emergent feature of the energy minimization, so that \mlm{it has not been possible to design a} specific desired critical configuration. \cds{One of our main results is that the space of critical configurations is a smooth manifold almost everywhere and consequently \mlm{we can explore disordered networks beyond those that are poised at an unprogrammed strain-induced critical point. Specifically, we derive an algorithm that finds critical networks with other optimized desired properties, such as maximizing the jump in bulk modulus or being composed of fibers of the same length.} These results shed new light on an under-explored problem in metamaterials: which material properties can be designed and to what degree?} \mlm{And in contrast to much previous work focused on ordered structures, such as miura-ori patterns in origami~\cite{silverberg_applied_2014}, our method explicitly encompasses a much broader set of disordered networks.}

\cds{Our result takes advantage of the distinction between ``first-order'' and ``second-order'' rigidity.}
\mlm{Indeed, models for both types of biological materials in the previous paragraph} -- vertex models for confluent tissues and spring networks for biopolymer networks -- can undergo rigidity transitions without any change to their network topology. \mlm{This is in stark contrast to the}
jamming transition in granular systems, \mlm{where the contact network topology changes as the density increases, and the material rigidifies when the number of contacts equals the number of degrees of freedom.} 
In this case, first-order perturbations to the lengths of the contacts constrain all possible infinitesimal motions of the system, and so the system is said to be ``first order rigid". 
\mlm{But in the examples of tissues and biopolymer networks, the number of constraints is often less that the number of degrees of freedom -- i.e. the systems are \emph{underconstrained}.} In those cases, rigidity occurs at the onset of geometric incompatibility \cite{merkel_minimal-length_2019}. These systems are called ``second order rigid" because they possess an extensive number of modes \mlm{that cost zero energy with respect to first-order perturbations to constraints, but do cost energy at second order.} 
\mlm{Although enumerating states near a first-order rigid jamming transition is notorious difficult due to the larger number of nearly ``kissing contacts"~\cite{donev_underconstrained_2007} and the related Gardner transition \cite{berthier_gardner_2019},} \cds{second-order rigidity occurs even with a fixed network topology and, consequently, depends only on its geometry.}

\mlm{As we show in more detail, the space of critical configurations
is naturally parameterized using the states of self-stress.}
This parameterization is \mlm{related to} the force density method, which has been used for form-finding of pre-tensioned nets \cite{schek_force_1974, malerba_extended_2012} and \cds{tensegrities} \cite{tran_advanced_2010, wang_form-finding_2021} in regular lattices.
\mlm{The parameterization} \cds{enables the ability to optimize a broad array of} objective functions \mlm{that can be written in terms of the states of self stress and therefore minimized via existing gradient descent algorithms.} \cds{We demonstrate this by} \cds{numerically generating and investigating optimal configurations for several objective functions.} \mlm{To find networks with enhanced elastic response at the transition, we maximize either the bulk modulus or shear modulus, and to obtain networks with tailored structure we minimize the fluctuations in the edges lengths or the edge tensions. We also compare these optimized configurations to networks taken from random samples of two different probability densities over the critical manifold. This provides insight into design rules for multifunctional mechanical metamaterials that could eventually guide manufacturing or self-assembly.} 



\section{Mathematical Preliminaries}

An important aspect of any mechanical system is its rigidity: are there deformations that cost zero energy? Traditionally this question is described in terms of the degrees of freedom of a system and a set of constraints $f_\alpha$, each of which arise from a quadratic term in the energy
\begin{align}
    E = \frac{1}{2}\sum_{\alpha} f_\alpha^2. \label{EnergyConstraints}
\end{align}
For example, consider a network of $N_b$ Hookean springs connecting $N_v$ vertices with the energy functional $E = \sum_{\alpha} (L_\alpha - \ell_\alpha)^2/2$ where $L_\alpha$ and $\ell_\alpha$ are the actual and prescribed lengths of edge $\alpha=1,...,N_b$. The degrees of freedom are the positions of the vertices $x_{i\mu}$, where $i=1,...,N_v$ indexes the vertices and $\mu = 1,..,d$ indexes the spatial dimensions. Each edge contributes one constraint, $f_\alpha = L_\alpha - \ell_\alpha$. 

In some cases, one can simply linearize the system and consider whether any infinitesimal motions are permitted by the constraints. This information is contained in the rigidity matrix $R_{\alpha i\mu} = \partial f_\alpha / \partial x_{i\mu}$, which relates a small perturbation to the degrees of freedom $\delta x_{i\mu}$ with the resulting first-order change to the constraints:
\begin{align}
    \delta f_\alpha = \sum_{i\mu} \frac{\partial f_\alpha}{\partial x_{i\mu}}\delta x_{i\mu} = \sum_{i\mu} R_{\alpha i\mu} \delta x_{i\mu} = 0. \label{FirstOrder}
\end{align}
The rigidity matrix specialized to Hookean springs, $\tilde{R}$, can be written as
\begin{align}
    \tilde{R}_{\alpha i\mu} = \frac{\partial f_\alpha}{\partial x_{i\mu}} = g_{\alpha i}\frac{L_{\alpha\mu}}{L_\alpha}, \label{RigidityMatrix}
\end{align}
where $g_{\alpha i}$ is the signed incidence matrix of an underlying directed graph representing the network that describes which vertices are connected by each edge, and $L_{\alpha\mu}$ is the $\mu$ component of the vector along edge $\alpha$. These bond vectors can in turn be written in terms of the vertex coordinates as 
\begin{align}
    L_{\alpha\mu} = \sum_i g_{\alpha i}x_{i\mu} + b_{\alpha\mu}, \label{BondVectors}
\end{align}
where $b_{\alpha\mu}$ is a constant matrix that encodes the boundary conditions of the network, which can come from being in a periodic box or from having a subset of vertices pinned in place. Additional details on how to derive $b_{\alpha\mu}$ can be found in the Supplemental Materials.

Vectors in the right nullspace of $R$ are called linear zero modes (LZM's) as they result in no first-order changes to the constraints. $R$ also relates the tensions on the edges of the network to the resulting forces on the vertices $F_{i\mu}$:
\begin{align}
    -F_{i\mu} = \frac{\partial E}{\partial x_{i\mu}} = \sum_\alpha \frac{\partial E}{\partial f_\alpha}\frac{\partial f_\alpha}{\partial x_{i\mu}} = \sum_\alpha f_\alpha {R}_{\alpha i\mu}. \label{Equilibrium}
\end{align}

If the constraints are not all satisfied $(f_\alpha \neq 0)$ we say the network is prestressed, and Eq. (\ref{Equilibrium}) implies that $f_\alpha$ must be in the left nullspace of $R$ in order for the network to be in equilibrium. For a given configuration, any vector $\bm\sigma$ in the left nullspace of $R$ 
\begin{align}
\sum_\alpha \sigma_\alpha R_{\alpha i\mu}, \label{selfstressdef}
\end{align}
which is called a state of self-stress, corresponds with a possible set of tensions on the edges of the network that result in force balance on the vertices. Note that the self-stresses are purely geometric objects that merely describe the possible ways that a given configuration can support internal forces.

The rank-nullity theorem applied to the rigidity matrix results in Maxwell-Calladine constraint counting, which relates the number of independent LZM's ($N_0$) and self-stresses ($N_s$) to the the number of degrees of freedom ($dN_v$) and constraints ($N_b$):
\begin{align}
    N_0 - N_s = dN_v - N_b. \label{Maxwell-Calladine}
\end{align}

If a system has no linear zero modes (excluding trivial rigid-body translations and rotations), then it is called first-order or infinitesimally rigid. Systems that undergo a jamming transition are often first-order rigid: as the free volume decreases, the number of contact between particles (and hence the number of constraints) increases until all LZM's are eliminated. 

However, underconstrained systems can also become rigid despite having an extensize number of LZM's, when nonlinear effects cause the LZM's to change the constraints at higher order. For example, sub-isostatic spring/fiber networks and vertex/Voronoi models for confluent tissues undergo a strain/shape-controlled rigidity transition despite being underconstrained \cite{bi_density-independent_2015, merkel_geometrically_2018, merkel_minimal-length_2019}, and packings of non-spherical particles become jammed below the isostatic point \cite{donev_underconstrained_2007, hecke_jamming_2009}. In these systems the LZM's change the constraints at quadratic order, so they are said to be second-order rigid. 

Underconstrained systems in a generic configuration do not have any self-stresses. In this case, any linear zero mode can always be extended into a non-linear floppy mode \cite{connelly2022frameworks}. However, in certain configurations the system has at least one self-stress. We call these special states critical configurations, and the set of all critical configurations we call the critical manifold. The existence of a self-stress leads to an additional necessary (but not sufficient) condition that a LZM $\dot{x}_{i\mu}$ must satisfy in order to be extended into a non-linear mode \cite{connelly_rigidity_1980, connelly_second-order_1996, damavandi_energetic_2022}:
\begin{align}
    \sum_{\alpha ij} \sigma_\alpha^I \frac{\partial^2 f_\alpha}{\partial x_{i\mu} \partial x_{j\nu}}\dot{x}_{i\mu} \dot{x}_{j\nu} =  \dot{\bm{x}}^T P^I \dot{\bm{x}} = 0, \label{SecondOrderRigidity}
\end{align}
for all self-stresses $\sigma^I$. If no LZM's satisfy these conditions, the system is second-order rigid. In practice, many second-order rigid systems satisfy a weaker condition called prestress stability, where there is a particular self-stress $\sigma_\alpha$ for which the associated prestress matrix $P_{i\mu j\nu}= \sum_\alpha \sigma_\alpha (\partial^2 f_\alpha/\partial x_{i\mu} \partial x_{j\nu})$ is positive-definite on the subspace of linear zero modes, which clearly implies second-order rigidity. In this work, we study networks that have at most one independent self-stress, for which second-order rigidity and prestress stability are equivalent \cite{connelly_second-order_1996}.

\begin{figure*}
\centering
\includegraphics[scale= 1]{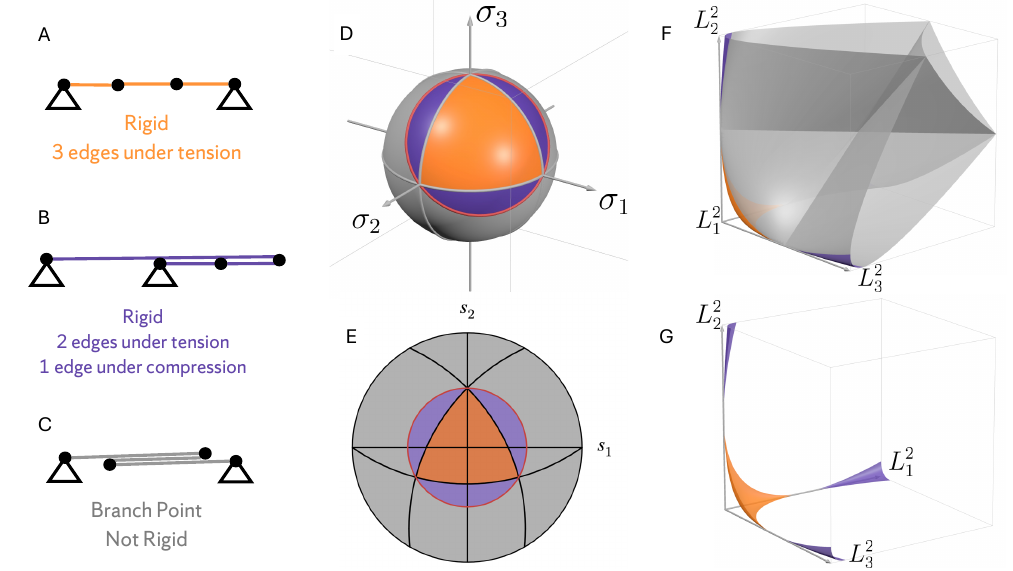}
\caption{\textbf{The critical manifold of the three bar linkage.} A,B) Examples of two different rigid critical configurations of the three bar linkage.
C) Example of a critical configuration that is a non-rigid branch point. D) All points on the critical manifold have a corresponding geometric stress. The sphere represents the space of these geometric stresses for the 3-bar linkage: the orange and purple regions contain the rigid
configurations, which look like A and B respectively; the grey region contains the branch points like C; the red line contains choices of
geometric stress for which the parameterization is not defined. As rescaling the geometric stress by the
same factor results in the same real-space configuration (Eq. \ref{SelfStressParamaterization_coords}), which is why the space of geometric stresses is an $N_b - 1$ -dimensional space,
here represented as a 2-sphere. E) A stereographic projection of the space of geometric stresses as described in Eq. (\ref{projection}). F) The critical manifold embedded in the space of squared lengths of the edges of the three bar linkage, where the different colored regions are the images of the corresponding regions of
geometric stresses. G) Same as panel F, except just the rigid parts of the critical manifold are highlighted for clarity.}
\label{3-barlinkage}
\end{figure*}

Underconstrained networks are generically floppy because the lengths of edges for which critical configurations can occur are a set of measure zero. Consider the three-bar linkage in Fig. \ref{3-barlinkage}, a simple example of an underconstrained system. In a generic configuration, the linkage has no self-stress, so the LZM corresponding with a shearing motion can always be extended into a non-linear zero mode. However, if all vertices are colinear, as shown in Fig \ref{3-barlinkage}A-C, a state of self-stress appears which can rigidify the linkage or induce a branch point. While these critical configurations are only a measure-zero subset of configuration space, the system can be driven to these special points due to geometric incompatibility. For example, if the distance between the ends of the linkage is larger than the sum of the preferred lengths of the three edges, the constraints cannot all be satisfied. This leads to inevitable prestress, and energy minimization then drives the system to a critical configuration. This mechanism of geometric incompatibility in the preferred geometry has been shown to be the cause of rigidity in general underconstrained spring networks as well as in 2- and 3-dimensional vertex and Voronoi models \cite{merkel_minimal-length_2019}. For the three bar linkage, we can visualize the rigid portion of the critical manifold as the boundary of geometric incompatibility. In Fig. \ref{3-barlinkage}F we show the full critical manifold of the three-bar linkage in the space of the squared lengths of the three edges \cite{berry_controlling_2022}. The grey surfaces correspond to branch points, and the orange and purple surfaces correspond with rigid configurations. The sub-surface of rigid configurations divides the space into two separate regions, which are shown in Fig. \ref{3-barlinkage}G. Sets of squared edge lengths that are longer than those on the colored manifold, corresponding to points in the upper right corner of Fig~\ref{3-barlinkage}G, are compatible with the boundary conditions, so choosing the rest lengths from this region will result in a floppy configuration like \ref{3-barlinkage}A. However, choices of squared edge lengths interior to the colored manifold in the lower left-hand corner of Fig \ref{3-barlinkage}G are impossible to achieve, and the resulting prestress then drives the system to a rigid critical configuration.

\section{Analytical Results}

\subsection{Geometric Stress Parameterization}
When reaching the second-order rigidity transition by straining a network to the point of geometric incompatibility, the resulting self-stress is an emergent feature of the energy minimization, and there is no clear way to predict the properties of the resulting rigid network. In the rest of this work, we describe a method that characterizes the entire critical manifold of any network with a given graph structure and boundary conditions using a natural parameterization that can be used to rationally explore the critical manifold and obtain configurations with specific properties.

Let us consider a network with an arbitrary energy functional $E(x_{i\mu}, p_n)$ that depends on the node coordinates and any other control parameters $p_n$, such as rest lengths or stiffnesses. Any equilibrium configuration of this network must satisfy force balance: 
\begin{align}
    \frac{\partial E}{\partial x_{i\mu}} = \sum_\alpha \frac{\partial E}{\partial h_\alpha}\frac{\partial h_\alpha}{\partial x_{i\mu}} = 0,
    \label{ForceBalance}
\end{align}
where $h_\alpha(x_{i\mu})$ can be any set of functions of the coordinates that can be used to completely determine the energy: $E(h_\alpha(x_{i\mu}))$. Here $h_\alpha$ plays a similar role to the constraint $f_\alpha$ in the previous section; however, while $f_\alpha$ is completely determined by the specific energy function being used, $h_\alpha$ can be chosen to be a more general coarse-graining of the node coordinates into higher-order geometric quantities that determine the energy. This allows us to write Eq. (\ref{ForceBalance}) separately in terms of the conjugate variables $\partial E /\partial h_\alpha$, which are the generalized tensions associated with the $h_\alpha$'s and contain all information about $E$, and $\partial h_\alpha /\partial x_{i\mu}$, which plays the role of the rigidity matrix by connecting the higher-level geometry of the $h_\alpha$'s to the lower-level geometry of the vertex coordinates.\mlm{We term this generalized rigidity matrix $ R^h_{\alpha i\mu} = \partial h_\alpha /\partial x_{i\mu}$ the ``geometric response matrix".}

We can then find critical configurations by looking for states that satisfy the force balance of Eq. (\ref{ForceBalance}) despite having non-zero internal stresses $\partial E /\partial h_\alpha \neq 0$. To do this, we will use the generalized tensions as a different set degrees of freedom, defining $\sigma^h_\alpha = \partial E /\partial h_\alpha$. \mlm{We call $\sigma^h_\alpha$ the ``generalized states of self-stress" corresponding to the geometric response matrix for a given $h$.} Then the force balance condition $\sum_\alpha \sigma^h_\alpha (\partial h_\alpha /\partial x_{i\mu}) = 0$ can be thought of as implicitly defining the critical configuration $x_{i\mu}(\sigma^h_\alpha)$ corresponding to each choice of generalized tensions. \mlm{For brevity, in remainder of this paper we sometimes refer to these as simply "states of self-stress", though they are distinct from the standard $\sigma$ defined in Eq.~\ref{selfstressdef} when $h_\alpha \neq f_{\alpha}$.}

However, most choices of $h_\alpha$ do not result in conjugate variables that generate a proper set of degrees of freedom. To see this, let us now assume that we have a central force network whose energy functional depends only on the edge lengths: $E(L_\alpha, p_n)$. One might think to use $h_\alpha = L_\alpha$. In this case, the conjugate variables $\partial E /\partial L_\alpha = \tau_\alpha$ are just the actual tensions on the edges, and the force balance condition is 
\begin{align}
    \frac{\partial E}{\partial x_{i\mu}} = \sum_\alpha  \frac{\partial E}{\partial L_\alpha}\frac{\partial L_\alpha}{\partial x_{i\mu}} = \sum_\alpha \tau_\alpha \left(g_{\alpha i}\frac{L_{\alpha\mu}}{L_\alpha} \right) = 0.
    \label{ForceBalanceL}
\end{align}
For a fixed $\tau_\alpha$, Eq. (\ref{ForceBalanceL}) cannot be solved to get a corresponding configuration $x_{i\mu}$. Consider the three-bar linkage in Fig. \ref{3-barlinkage}A; any such critical configuration with all edges under tension must have equal tension on all edges to maintain force balance: $\tau_1 = \tau_2 = \tau_3$. Therefore, if we choose $\tau_\alpha \propto [1\, 1\, 1]^T$ Eq. (\ref{ForceBalanceL}) will have infinitely many solutions, and for any other choice of $\tau_\alpha$ Eq. (\ref{ForceBalanceL}) will have no solutions. Therefore we reject this choice of $h_\alpha$ and its conjugate self-stress $\tau_\alpha$.

Instead, to ensure we can solve the force balance equation, we should choose $h_\alpha$ to be the simplest function of $L_\alpha$ that is quadratic in $x_{i\mu}$ so that taking a derivative results in an equation that is linear in the coordinates. We can do this using $h_\alpha = L_\alpha^2/2$, which results in the conjugate variable
\begin{align}
    \sigma^h_\alpha = \frac{\partial E}{\partial (L_\alpha^2/2)} = \frac{\partial E}{\partial L_\alpha}\frac{1}{L_\alpha} = \frac{\tau_\alpha}{L_\alpha},\label{ForceDensity}
\end{align}
which corresponds with the force density (tension per length) of each edge. We use this choice of $h_\alpha$ and the corresponding generalized self-stress for the remainder of this work. We will refer to this particular self-stress as the geometric stress, as we demonstrate below that it naturally encodes the geometry of the critical manifold. The force balance condition then becomes
\begin{align}
    \frac{\partial E}{\partial x_{i\mu}} &= \sum_\alpha \frac{\partial E}{\partial (L_\alpha^2/2)}\frac{\partial (L_\alpha^2/2)}{\partial x_{i\mu}} \nonumber\\
    &= \sum_\alpha \sigma^h_\alpha \left(g_{\alpha i}L_{\alpha\mu} \right) = 0,
    \label{ForceBalanceLsqr}
\end{align}
which eliminates the factor of $1/L_\alpha$ in Eq. (\ref{ForceBalanceL}). Now, for any choice of $\sigma^h_\alpha$, we can write Eq. (\ref{ForceBalanceLsqr}) as a linear equation in the coordinates:
\begin{align}
    \frac{\partial E}{\partial x_{i\mu}} &= \sum_\alpha \sigma^h_{\alpha} g_{\alpha i} L_{\alpha\mu}= \sum_{\alpha j} \sigma^h_{\alpha} g_{\alpha i}(g_{\alpha j}x_{j\mu} + b_{\alpha\mu}) \\
    &= \sum_{j\nu} \left(\sum_\alpha \sigma^h_{\alpha} g_{\alpha i}g_{\alpha j}\delta_{\mu\nu}\right) x_{j\nu} + \sum_\alpha \sigma^h_{\alpha} g_{\alpha i}b_{\alpha\mu} \\
    &= \sum_{j\nu}P_{i\mu j\nu}\,x_{j\nu} + B_{i\mu} = 0,
    \label{FDM1}
\end{align}
where $P$ plays the role of the prestress matrix, and is given by
\begin{align}
    P_{i\mu j\nu} = \sum_{\alpha} \sigma^h_\alpha \frac{\partial^2 h_\alpha}{\partial x_{i\mu} \partial x_{j\nu}} = \sum_{\alpha} \sigma^h_\alpha g_{\alpha i}g_{\alpha j} \delta_{\mu\nu}.  \label{PrestressMatrix}
\end{align}

Notice that this geometric prestress matrix has no explicit dependence on the coordinates $x_{i\mu}$, and depends only on the geometric stress $\bm\sigma^h$ and the structure of the underlying graph encoded by $g$. $P$ is also block-diagonal with the respect to the spatial directions ($P_{i\mu j\nu} = P_{ij}\delta_{\mu\nu}$), and the non-zero blocks are exactly the weighted Laplacian matrix associated with with the underlying graph with the geometric stress as the weights. We will sometimes refer to just this non-zero block as the prestress matrix, in which case it will have only two indices: $P_{ij}=\sum_{\alpha} \sigma^h_\alpha g_{\alpha i}g_{\alpha j}$.

This matrix always has the trivial zero mode $[1 1 \dots 1]^T$ corresponding to rigid-body translations. In this work we will choose the components $\sigma^h_\alpha$ to be all positive, corresponding to networks with all edges under tension. In this case, $P$ has no other zero modes, so Eq. (\ref{FDM1}) can always be solved to obtain a unique critical configuration up to translations:
\begin{align}
    x_{i\mu}(\bm\sigma^h) = -\sum_j \left(P^{-1}\right)_{ij}\left(\sum_\alpha \sigma^h_{\alpha} g_{\alpha j}b_{\alpha\mu}\right),
    \label{SelfStressParamaterization_coords}
\end{align}
where ${P}^{-1}$ should be understood as the pseudo-inverse of $P$. Note that the translational zero mode $[1 1 ... 1]^T$ is also in the nullspace of the incidence matrix $g_{\alpha j}$, which guarantees that Eq. (\ref{FDM1}) has a solution.

To describe each critical configuration uniquely, we can instead write the parameterization directly for the vectors along the edges:
\begin{align}
    L_{\alpha\mu}(\bm\sigma^h) = \sum_\beta\left\{\delta_{\alpha\beta} - \left[\sum_{ij}g_{\alpha j}\left({P}^{-1}\right)_{ij}g_{\beta j}\right]\sigma^h_{\beta}\right\}b_{\beta\mu}.
    \label{SelfStressParamaterization_bondvecs}
\end{align}

The mapping in Eq. (\ref{SelfStressParamaterization_coords}) provides a natural parameterization of the critical manifold: while a generic choice of vertex coordinates will not result in a critical configuration, almost any choice of $\sigma^h_\alpha$ will produce a unique critical configuration (up to rigid-body motions). This is essentially the force-density method, which has been used by engineers to find equilibrium shapes for tensioned nets and tensegrity structures that are stabilized by internal stresses \cite{schek_force_1974, wang_form-finding_2021}. Of course, for the special case where the energy functional represents springs with no rest length, $f_{\alpha} = L^2_{\alpha}/2$, the geometric response matrix $\partial h_\alpha /\partial x_{i\mu}$ is the standard rigidity matrix.

Note that this geometric stress parameterization can still be used for overconstrained networks, although it is significantly less useful: since constraint counting implies that every configuration possesses a self-stress, the critical manifold fills all of configuration space, negating the need for an alternate parameterization. In addition, for networks with multiple self-stresses, the mapping is no longer one-to-one, as multiple independent geometric stresses get mapped onto the same critical configuration. In future work, it may be interesting to use this method to find any relationship that may exist between different self-stresses of the same configuration. However, for this work we will focus only on networks with exactly one self-stress.

Once the configuration corresponding to a given $\bm\sigma^h$ has been obtained, we still need to choose any control parameters such that force balance is actually achieved with the desired energy functional. For example, if we want a network with a Hookean energy $E = \sum k_\alpha/2(L_\alpha-\ell_\alpha)^2$, we can ensure it is in mechanical equilibrium by choosing the rest lengths and stiffnesses that satisfy $\partial E/\partial (L_\alpha^2/2) = k_\alpha(L_\alpha - \ell_\alpha)/L_\alpha = c\sigma^h_\alpha$. Choosing $c=0$ (setting $\ell_\alpha = L_\alpha$) results in the configuration being at the critical point, as it possesses a self-stress but is not under any actual stress.


Different choices of the energy functional result in different effective stiffness of the resulting configuration, which (following \cite{guest_stiffness_2006}) we can see by writing the Hessian matrix of a generic $E(L_\alpha)$:
\begin{align}
    H_{i\mu j\nu} &= \frac{\partial^2 E}{\partial x_{i\mu}\partial x_{j\nu}}\nonumber{} \\
    &= \sum_{\alpha\beta} \frac{\partial h_\alpha}{\partial x_{i\mu}}\frac{\partial^2 E}{\partial h_\alpha \partial h_\beta}\frac{\partial h_\beta}{\partial x_{j\nu}} + \frac{\partial E}{\partial h_\alpha}\frac{\partial^2 h_\alpha}{\partial x_{i\mu}\partial x_{j\nu}}\nonumber{} \\
    &= {R^h}^T K R^h + P. \label{GenericHessian}
\end{align}

$K$ is a matrix of ``modified stiffnesses", which contains all information about the specific energy function: $R^h$ depends only on the geometry of the current configuration, and $P$ depends only on the current geometric stress. For a Hookean spring network, $K$ is a diagonal matrix with entries
\begin{align}
    K_{\alpha\alpha} = \frac{\partial^2 E}{\partial h_\alpha^2} = \frac{1}{L_\alpha^2} \left(\frac{\partial^2 E}{\partial L_\alpha^2} - \frac{1}{L_\alpha}\frac{\partial E}{\partial L_\alpha} \right) = \frac{k_\alpha - \sigma^h_\alpha}{L_\alpha^2}, \label{ModifiedStiffness}
\end{align}
where $k_\alpha = \partial^2 E/\partial L_\alpha^2$ is the actual stiffness of edge $\alpha$, which is constant for Hookean springs but could be a function of the edge length for a nonlinear spring potential. Other models can have non-zero entries off the main diagonal of $K$. For example, a vertex model without an area term would have $K_{\alpha\beta}\neq 0$ whenever edges $\alpha$ and $\beta$ are part of the same cell.

From Eq. (\ref{GenericHessian}), we can see that the for small amounts of prestress, the low-frequency spectrum of the Hessian is controlled solely by the prestress matrix, and is not affected by the specific energy functional. The linear zero modes, which are determined only by the geometry of the current configuration, are zero modes of $R^h$, and so are also zero modes of the first term in the Hessian. Therefore, their stiffnesses depend only on $P$, which in turn only depends on the geometric stress and the graph structure of the network. In the Supplemental Materials, we compare these expressions for the Gram and prestress terms of the Hessian to those derived from expanding the derivatives in terms of the constraints $f_\alpha = L_\alpha - \ell_\alpha$ rather than $h_\alpha = L_\alpha^2/2$.

\begin{figure}
\centering
\includegraphics[scale= 1]{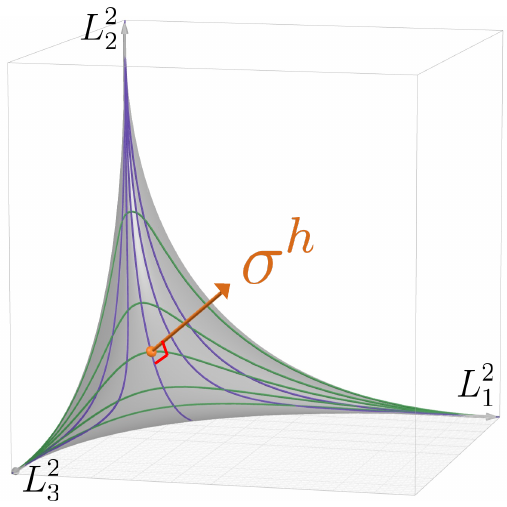}
\caption{\textbf{Geometry of the critical manifold of the three bar linkage.} Green/purple lines correspond with evenly spaced grid lines in the space of the Cartesian coordinates $s_1,s_2$ in Fig. \ref{3-barlinkage}E. For each point on the manifold, the geometric stress $\sigma^h$ (in orange) is always normal to the surface.}
\label{CriticalManifoldGeometry}
\end{figure}

\subsection{Geometry of the Critical Manifold}
Given that we have an analytic expression for the geometric stress parameterization, we can describe the geometry of the critical manifold, which we consider as a subset of the space of squared edge lengths,
\begin{align}
    h_{\alpha}(\bm\sigma^h) = \frac{1}{2}L^2_{\alpha}(\bm\sigma^h) = \frac{1}{2}\sum_\mu L^2_{\alpha\mu}(\bm\sigma^h).
    \label{SelfStressParamaterization_Lsqr}
\end{align}

This surface is a smooth manifold almost everywhere, except for choices of $\bm\sigma^h$ for which the prestress matrix has non-trivial zero modes and thus Eq. (\ref{FDM1}) has no unique solution (up to translations). As this space is parameterized by the $N_b$ components of the geometric stress, one might expect the critical manifold to be $N_b$-dimensional. However, the parameterization in Eq. (\ref{SelfStressParamaterization_bondvecs}) is invariant under rescaling the geometric stress: $L_{\alpha\mu}(c\bm\sigma^h) = L_{\alpha\mu}(\bm\sigma^h)$. This implies that we should think of the space of geometric stresses as the $(N_b-1)$-dimensional projective space $\mathcal{RP}^{N_b-1}$, for which the $N_b$ components $\sigma^h_\alpha$ form a set of homogeneous coordinates. In Fig \ref{3-barlinkage}D we represent this manifold for the 3-bar linkage as a 2-sphere. We can define a unique coordinate for each point on the manifold by taking a stereographic projection of the geometric stresses. We can choose a given unit-normalized geometric stress $\hat{\bm\sigma}^0$ as the ``north pole" of the projection and choose an orthonormal basis $\{\hat{\bm\sigma}^0 , \bm{v}^1 , \dots , \bm{v}^{N_b-1}\}$ for $\mathcal{R}^{N_b}$. Then we can specify each geometric stress by $N_b-1$ Cartesian coordinates $s_\gamma$: 
\begin{align}
\sigma^h_\alpha(s_\gamma) = \frac{\sum_\gamma v_{\alpha}^\gamma s_\gamma + (1-|\bm{s}|^2/4)\hat{\sigma}_\alpha^0}{1+|\bm{s}|^2/4}.\label{projection}
\end{align}
In Figure 1E we show this projection for the three bar linkage, using $\hat{\bm\sigma}^0 = (1/\sqrt{3})[1\, 1\, 1]^T$. When writing geometric quantities we will use the derivative of this projection
\begin{align}
V_{\alpha\gamma} := \frac{\partial \sigma^h_\alpha}{\partial s_\gamma} = \frac{v_\alpha^\gamma - s_\gamma[\hat{\sigma}_\alpha^0 + \sigma^h_\alpha(\bm{s})]/2}{1+|\bm{s}|^2/4}.\label{projectionDeriv}
\end{align}
While it is sometimes easier to use the projected coordinates $s_\gamma$ to write some geometric quantities (as we do below), we will typically just use the redundant homogeneous coordinates $\sigma^h_\alpha$.

Interestingly, in addition to acting as the coordinates for the critical manifold, the geometric stress $\bm\sigma^h$ is also normal to the surface \cite{berry_controlling_2022}. Consider any smooth trajectory $\sigma^h_\alpha(t)$ through the space of geometric stresses, and its image in the space of lengths $h_\alpha(\sigma^h(t))$. The tangent vectors along this curve are given by 
\begin{align}
    \frac{\partial h_\alpha}{\partial t} = \sum_{i \mu\beta}\frac{\partial h_\alpha}{\partial x_{i\mu}}\frac{\partial x_{i\mu}}{\partial \sigma^h_\beta}\frac{\partial \sigma^h_\beta}{\partial t}.
    \label{TangentVectors}
\end{align}
Then because the geometric stresses are defined as being in the nullspace of the geometric response matrix,
\begin{align}
    \sum_\alpha \sigma^h_\alpha \frac{\partial h_\alpha}{\partial x_{i\mu}}\Bigr|_{\sigma^h_\alpha} = 0,
\end{align}
the geometric stresses are orthogonal to the tangent vectors at each point on the critical manifold
\begin{align}
    \sum_\alpha \sigma^h_\alpha \frac{\partial h_\alpha}{\partial t}\Bigr|_{\sigma^h_\alpha} = 0,
\end{align}
so the geometric stresses are normal to the surface.

We want to be able to search the critical manifold for particular configurations. To do this, we need to know how a small step in geometric stress space changes the resulting critical configuration: $\delta L_{\alpha\mu} \sim (\partial L_{\alpha\mu}/\partial\sigma^h_\beta) \delta \sigma^h_\beta$. We show in the Supplemental Materials that this derivative of the parameterization with respect to the geometric stress is given by
\begin{align}
    \frac{\partial L_{\alpha\mu}}{\partial \sigma^h_\beta} = -\left[\sum_{ij}g_{\alpha i}\left(P^{-1}\right)_{ij} g_{\beta j}\right]L_{\beta\mu},
    \label{SelfStressParamaterization_BondVectorDeriv}
\end{align}
which we can then use to write the derivatives of the squared lengths:
\begin{align}
    \frac{\partial h_{\alpha}}{\partial \sigma^h_\beta} &= \sum_{\mu}L_{\alpha\mu}\frac{\partial L_{\alpha\mu}}{\partial \sigma^h_\beta} = -\sum_{ij\mu} L_{\alpha\mu}g_{\alpha i}\left(P^{-1}\right)_{ij}g_{\beta j}L_{\beta\mu} \nonumber\\
    &= -\left(R^h P^{-1} {R^h}^T\right)_{\alpha\beta} := B_{\alpha\beta},
    \label{SelfStressParamaterization_LsqrDeriv}
\end{align}

This is a key equation that allows us to search the geometric stress space in order to optimize material properties that are naturally written as a function of $L_{\alpha\mu}$, as shown in Eq. (\ref{TotalDeriv}) and discussed in the next section.

Notice that this matrix is symmetric, because it is closely related to the bond operator which has been used to quantify the elastic response of amorphous particle packings \cite{giannini_bond-space_2021, lerner_breakdown_2014}:
\begin{align}
    \Xi_{\alpha\beta} = \left(RH^{-1}R^T\right)_{\alpha\beta},
    \label{BondOperator}
\end{align}
which gives the change in length of bond $\alpha$ due to a force dipole on bond $\beta$. Here $H$ is the full Hessian matrix and $R=\partial L_\alpha/\partial x_i$ is the usual rigidity matrix. Since Eq. (\ref{SelfStressParamaterization_LsqrDeriv}) is a function of only the geometric prestress matrix rather than the full Hessian, it does not account for the full linear response of the network, but just describes how the structure must change to accommodate an external force dipole \emph{before} relaxation.

The matrix $B$ in Eq. (\ref{SelfStressParamaterization_LsqrDeriv}) also gives the tangent space at each point on the critical manifold. The columns of Eq. (\ref{SelfStressParamaterization_LsqrDeriv}) span the tangent space, but there is one redundant column because we are using the homogeneous coordinates $\sigma^h_\alpha$. We can get rid of this by using a projection as in Eq. (\ref{projection}) to write the $N_b-1$ tangent vectors:
\begin{align}
    T_\alpha^\gamma = \frac{\partial h_{\alpha}}{\partial s_\gamma} = \sum_\beta \frac{\partial h_{\alpha}}{\partial \sigma^h_\beta}\frac{\partial \sigma^h_\beta}{\partial s_\gamma} = (BV)_{\alpha\gamma}.
    \label{TangentSpace}
\end{align}

We can then write the metric as the inner products of the tangent vectors:
\begin{align}
    g_{\gamma\delta} = \sum_\alpha T_\alpha^\gamma T_\alpha^\delta = (V^T B^2 V)_{\gamma\delta}.
    \label{Metric}
\end{align}

The fact that the geometric stress is also the normal vector to the smooth part of the critical manifold results in an interesting coincidence. We show in the Supplemental Materials that the second fundamental form of the critical manifold, which describes the local curvature information, is given by
\begin{align}
    \mathrm{I\!I}_{\gamma\delta} = \sum_\alpha \frac{\partial^2 h_\alpha}{\partial s_\gamma \partial s_\delta}\frac{\sigma^h_\alpha}{|\bm\sigma^h|} = \frac{1}{|\bm\sigma^h|} (V^T B V)_{\gamma\delta}.
    \label{2FF}
\end{align}

This means essentially all geometric information about the critical manifold is contained in the matrix $B$ from Eq. (\ref{SelfStressParamaterization_LsqrDeriv}), along with whatever projection is chosen from the homogeneous coordinates $\sigma^h_\alpha$ to the Cartesian coordinates $s_\gamma$.


\section{Numerical Results}

\begin{figure}
\centering
\includegraphics[scale= 1]{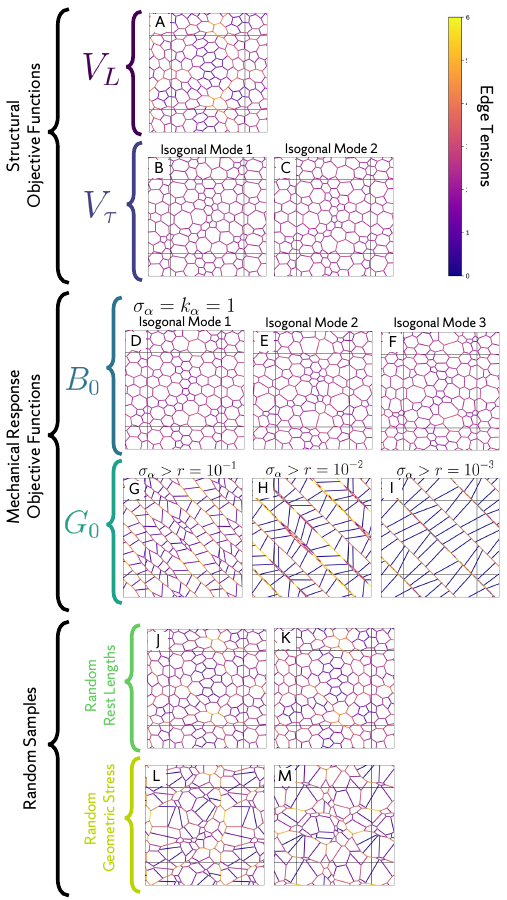}
\caption{\tah{\textbf{Representatives of all groups of optimized and random critical configurations from a single network structure.} A) The configuration that minimizes the length fluctuations $V_L$. B,C) Two degenerate optimal configurations of the tension fluctuations $V_\tau$ that are related by an isogonal transformation. D) The configuration that maximizes the bulk modulus at the transition $B_0$ obtained by setting the geometric stress equal to the edge stiffnesses $(\sigma^h_\alpha = k_\alpha = 1)$. E,F) Two other configurations with optimum $B_0$ that are related to the configuration in D) by two different isogonal modes. G,H,I) Configurations that maximize the shear modulus at the transition with various lower thresholds on the geometric stress. Note that lowering the threshold allows the network to become more aligned, increasing the resulting shear modulus. J,K) Two configurations generated by randomly assigning rest lengths and volumetrically straining to the critical point. L,M) Two configurations generated by sampling a random positive-definite geometric stress and using the geometric stress parameterization to obtain the corresponding configuration. As the tensions in the networks are defined only up to an overall scale, we fix the normalization of the geometric stress by setting the pressure equal to unity $(P = \sum_\alpha \sigma^h_\alpha L_\alpha^2/2V = 1)$ and coloring the edges by the resulting tensions $\tau_\alpha = \sigma^h_\alpha L_\alpha$.}}
\label{Networks}
\end{figure}

Here, we specialize the results from the previous section to two-dimensional periodic random regular networks generated from a Voronoi tessellation, which ensures that the networks are random regular graphs with a uniform connectivity $z=3$ and thus possess at most one self-stress. To keep the lengths of the edges $\mathcal{O}(1)$, we scale the size of the periodic box with the number of vertices: $V = N_v$. We also limit our networks to have positive-definite geometric stresses, equivalent to all edges being under tension rather than compression, although other possibilities would be interesting for future work. We assume all networks have a Hookean energy function 
\begin{align}
    E = \frac{1}{2}\sum_\alpha k_\alpha(L_\alpha - \ell_\alpha)^2, \label{HookeanEnergy}
\end{align}
where the edge stiffnesses $k_\alpha$ are set to unity unless stated otherwise.

Using the derivative of the geometric stress parameterization in Eq. (\ref{SelfStressParamaterization_BondVectorDeriv}), we can calculate the total derivative of any function $O(\sigma^h_\alpha, L_{\alpha\mu})$ with respect to the geometric stress:
\begin{align}
    \frac{\text{d} O}{\text{d} \sigma^h_\alpha} = \frac{\partial O}{\partial \sigma^h_\alpha} + \sum_{\beta\mu}\frac{\partial O}{\partial L_{\beta\mu}}\frac{\partial L_{\beta\mu}}{\partial \sigma^h_\alpha}.
    \label{TotalDeriv}
\end{align}

We then use the FIRE gradient descent algorithm \cite{guenole_assessment_2020} to optimize the objective function on the critical manifold (details in the Supplemental Materials). We consider two different types of objective functions: structure-based functions that describe the geometry of a network or its dual force network, and response-based functions that quantify the network's mechanical behavior.

\subsection{Generating Critical Networks}

We generate three classes of networks on the second order rigidity critical manifold.
\begin{enumerate}
    \item Response objective functions: networks optimized to maximize a specific mechanical response
    \item Structural objective functions: networks that optimize a specific structural quantity, and
    \item Random samples: networks that are constructed from random samples of a distribution over the critical manifold.
\end{enumerate}

\subsubsection{Response Objective Functions}

Underconstrained networks on the floppy side of the critical manifold generically have all of their elastic moduli vanish, but upon reaching the critical manifold the moduli jump discontinuously to a finite value. In the Supplemental Materials, we use linear response (following \cite{maloney_amorphous_2006}) to calculate this discontinuity in the elastic modulus associated with a bulk affine deformation parameterized by a strain $\gamma$:
\begin{align}
    \mu = \frac{1}{V}\frac{\text{d}^2 E}{\text{d}\gamma^2} &= \frac{1}{2V} \frac{\left(\sum_\alpha \sigma^h_\alpha \dot{h}_\alpha \right)^2}{\sum_\alpha (\sigma^h_\alpha)^2 h_\alpha / k_\alpha}, \label{GenericModulus}
\end{align}
where $\dot{h}_\alpha = \partial h_\alpha/\partial \gamma$ quantifies how the critical manifold changes due to the bulk deformation to the periodic box.

Past the critical manifold, the presence of prestress leads to additional corrections to the modulus, but right at the transition, where the prestress vanishes, we can write the modulus as a simple function of the geometric stress, squared lengths, and edge stiffnesses. Note that the geometric stress $\sigma^h_\alpha$ in Eq. (\ref{GenericModulus}) is explicitly acting as the coordinates for the critical manifold and not describing any actual tension on the edges (hence why $\mu$ is independent of the normalization of $\bm\sigma^h$).

Eq. (\ref{GenericModulus}) has a nice interpretation in terms of the geometry of the critical manifold. The numerator of Eq. (\ref{GenericModulus}) is maximized at the particular point on the critical manifold (that is, the particular critical configuration) where the bulk deformation causes the critical manifold to move parallel to the geometric stress, which is normal to the critical manifold. Then this configuration would be pushed furthest away from the critical manifold by a given strain. However, this argument does not account for the non-affine relaxation that occurs during a bulk deformation. If it costs a lot of energy to strain a configuration away from the critical manifold along the normal direction, then there will be a more non-affine deformation that decreases the elastic modulus. The denominator of Eq. (\ref{GenericModulus}) quantifies this energy cost. Since moving normal to the critical manifold has the effect of simply scaling the tensions, we can write
\begin{align}
    \frac{\partial E}{\partial h_\alpha} = k_\alpha\frac{L_\alpha - \ell_\alpha}{L_\alpha} = c\sigma^h_\alpha, 
\end{align}
where $c>0$ quantifies how far we have moved away from the manifold. Then we can write
\begin{align}
    \sum_\alpha (\sigma^h_\alpha)^2 h_\alpha / k_\alpha &= \frac{1}{c^2}\sum_\alpha k^2_\alpha\frac{(L_\alpha - \ell_\alpha)^2}{L^2_\alpha}\frac{L_\alpha^2}{2k_\alpha} \nonumber\\
    &= \frac{1}{c^2}\sum_\alpha \frac{k_\alpha}{2} (L_\alpha - \ell_\alpha)^2 = \frac{E(c)}{c^2}.
\end{align}

Since the energy grows quadratically with $c$, this quantity is constant and finite even as $c$ goes to 0. Therefore, we can think of the process of maximizing Eq. (\ref{GenericModulus}) as looking for a configuration that can be moved the largest distance away from the critical manifold for the smallest energy cost.

First, we optimize the bulk modulus at the rigidity transition, which is given by
\begin{align}
    B_0 &= V\frac{\text{d}^2 E}{\text{d}V^2} = \frac{1}{4V}\frac{\left(\sum_\alpha \sigma^h_\alpha L_\alpha^2 \right)^2}{\sum_\alpha (\sigma^h_\alpha)^2 L_\alpha^2}, \label{BulkModulus}
\end{align}
since $\dot{h}_\alpha = \partial h_\alpha/\partial V = h_\alpha/V$, and where for we have taken the stiffnesses of the edges $k_\alpha = 1$ as the natural unit.

As with the structural objective functions, we consider this quantity as a function only of the geometric stress, since the geometric stress parameterization gives us $L_\alpha(\bm\sigma^h)$. We then calculate the total derivative $\text{d} B_0/\text{d} \sigma^h_\alpha$ and use gradient ascent to find a critical configuration that maximizes $B_0$.

We find that there is a degenerate subset of critical configurations that have the same optimum value of $B_0$. 
All degenerate optima of $B_0$ possess the same dual force network, and thus are related by isogonal modes \cite{noll_active_2017, brauns_geometric_2024}. To \mlm{demonstrate} this, we calculate the Hessian matrix of $B_0$ with respect to the geometric stress $B_0'' = \partial^2 B_0/\partial\sigma^h_\alpha \partial\sigma^h_\beta$ in order to quantify the shape of the cost landscape. We find that this matrix is negative semi-definite, with the number of flat directions equal to the number of polygons in the network.
\begin{align}
    \text{Null}(B_0'') = N_v/2 - 1 = N_f. \label{BulkModulusHessian}
\end{align}
Since any isogonal mode can be written as a sum of modes that expand or contract the individual polygons of the network, this counting argument is consistent with the degenerate subspace of optima possessing the same force network. 
$B_0$ cannot be written as a function only of the tensions, \mlm{so this is a somewhat surprising symmetry that should be studied in future work.}

\mlm{To parameterize the degenerate subset of configurations with the same bulk modulus $B_0$, we write the modulus as}
a sum of affine and non-affine contributions
\begin{align}
    B_0 &= B_0^{\text{aff}} + (B_0 - B_0^{\text{aff}}), \label{BulkModulusAffNonAff}
\end{align}
where $B_0^{\text{aff}} = V\frac{\partial^2 E}{\partial V^2} = \sum_\alpha k_\alpha L_\alpha^2 /4V$.
\mlm{Configurations in the degenerate subset} differ in how $B_0$ is split between the affine and non-affine terms. Interestingly, the most affine configuration, with non-affine component equal to zero, is given by
\begin{align}
    0 &= B_0 - B_0^{\text{aff}} = \frac{1}{4V}\frac{\left(\sum_\alpha \sigma^h_\alpha L_\alpha^2 \right)^2}{\sum_\alpha (\sigma^h_\alpha)^2 L_\alpha^2/k_\alpha} - \sum_\alpha k_\alpha L_\alpha^2 /4V , \label{BulkModulusAffine}
\end{align}
\mlm{which occurs precisely when the geometric stress is proportional to the the edge stiffness,}  $\sigma^h_\alpha = k_\alpha$. \mlm{The most affine example shown in Fig.~\ref{Networks}D, and several other isogonal modes are shown in Fig.~\ref{Networks}E,F. One can think of this as an algorithm to construct the degenerate space of maximal $B_0$ configurations: they are the set of all configurations with the same force network as the state where $\sigma^h_\alpha = k_\alpha$.}


We also attempt to maximize the shear modulus of a network at the rigidity transition. While the shear modulus of an isotropic second-order rigid network vanishes at the transition point in the thermodynamic limit, anisotropic networks can have a finite shear modulus. We choose the modulus associated with the pure shear deformation $$A_{\mu\nu} = I + \gamma \left[\begin{array}{cc}
    0 & 1/2 \\
    1/2 & 0
\end{array}\right],$$
which is given by 
\begin{align}
    G_0 &= \frac{1}{V}\frac{\text{d}^2 E}{\text{d}\gamma^2} = \frac{1}{V}\frac{\left(\sum_\alpha \sigma^h_\alpha L_{\alpha x}L_{\alpha y} \right)^2}{\sum_\alpha (\sigma^h_\alpha)^2 L_\alpha^2}, \label{ShearModulus}
\end{align}
where we have again set the edge stiffnesses to $k_\alpha=1$.

\begin{figure*}
\centering
\includegraphics[scale= 1]{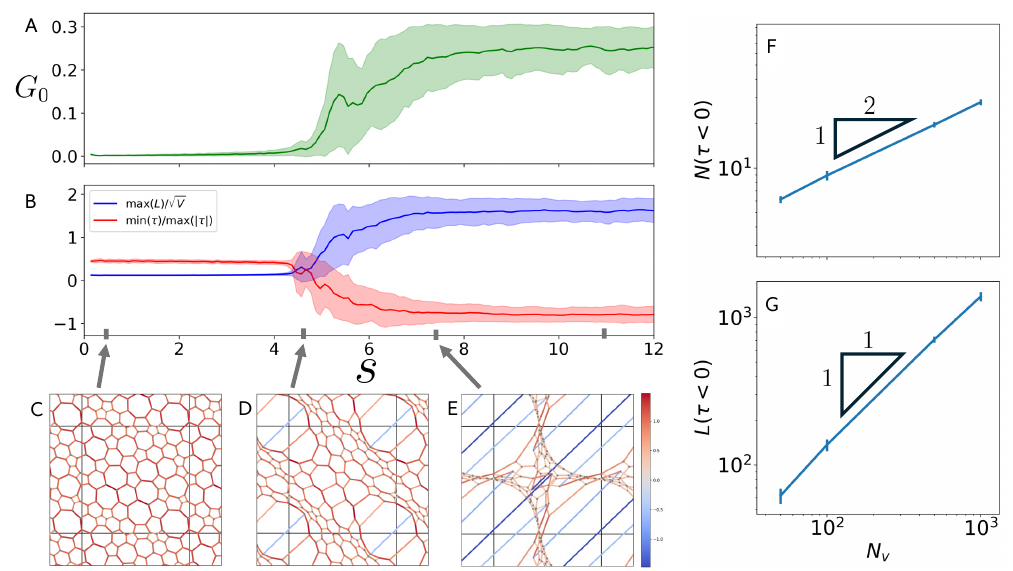}
\caption{\textbf{Unconstrained optimization of the shear modulus $G_0$} A) Shear modulus as a function of effective number of optimization steps $s$ during the optimization procedure without any additional constraints on $\sigma^h_\alpha$, averaged over an ensemble of 100 networks. B) For the same ensemble, the length of the longest edge of the network in units of the linear box dimension $\sqrt{V}$ (blue) and the minimum component of the edge tensions $\tau_\alpha = L_\alpha \sigma^h_\alpha$ normalized by the largest-magnitude component of tension. C-E) Images of a network at different points along the optimization. Red edges are under tension, blue edges are under compression. See Movie 1 in the Supplemental Materials for an animation of the optimization. (F,G) Features of the final optimized networks as a function of system size: F) System size scaling of the number of edges under compression $N(\tau<0)$. G) System size scaling of the sum of lengths of edges under compression $L(\tau<0)$.}
\label{ShearModulusSteepestDescent}
\end{figure*}

However, this process is not as well-behaved as optimizing the bulk modulus, which results in a rigid configuration with a positive-definite geometric stress and edge lengths clustered around the natural length scale $\bar{L}=1$. In contrast, when walking through the space of geometric stresses along the direction of increasing shear modulus, a subset of edges that are aligned with the imposed shear direction begin to grow until they become on the order of the linear box size $\sqrt{N_v}$. In Fig. \ref{ShearModulusSteepestDescent}A-B we plot the shear modulus and two structural indicators during the optimization procedure for an ensemble of 100 networks. The colored regions corresponds with standard deviations over the ensemble, and the fluctuations are due to an adaptive time step in the gradient descent algorithm. In Fig. \ref{ShearModulusSteepestDescent}B we plot the length of the longest edge in the network normalized by the linear box dimension $\text{max}(L_\alpha)/\sqrt{N_v}$ in blue, and the minimum tension in the network normalized by the largest-magnitude component of tension $\text{min}(\tau_\alpha)/\text{max}(|\tau_\alpha|)$ in red (we normalize the tensions because, like $\bm\sigma^h$, they are defined only up to an overall scale factor). Then if $\text{max}(L_\alpha)/\sqrt{N_v}=1$ the longest edge spans the entire system, and if $\text{min}(\tau_\alpha)/\text{max}(|\tau_\alpha|) = -1$ the edge with the largest magnitude of tension is under compression. 

Fig. \ref{ShearModulusSteepestDescent}C-E show a network at various points along the optimization, where red edges are under tension and blue edges are under compression. The final configurations reached after the optimization are essentially tensegrity structures, which in two dimensions require a significant number of edges to wrap around the periodic box and cross over each other. For this reason we do not show these final configurations here, see Movie 1 in the Supplemental Materials for a full animation of a typical optimization. In Fig. \ref{ShearModulusSteepestDescent}G-I we investigate the size scaling of the subset of compressive edges. The number of edges with negative tension scales with the linear box size $N(\tau<0) \propto \sqrt{N_v}$, while the sum of lengths of negative-tension edges scales with the system area $L(\tau<0) \propto N_v$. This implies that the average length of the negative-tension edges $L(\tau<0)/N(\tau<0)$ also scales with the linear box size. 

This makes sense in the context of our interpretation of the numerator and denominator of the elastic moduli. We saw that any modulus becomes larger when the associated deformation of the critical manifold $\dot{\bm{h}}$ is close to being aligned with the normal vector $\bm\sigma^h$. For a volumetric expansion, $\dot{h}_\alpha = L_\alpha^2/2V$ is always positive definite, so it makes sense that the resulting optimal geometric stresses are also positive. However, for pure shear along the direction $45 \degree$ from the $x$-axis of the periodic box we have $\dot{h}_\alpha = L_{\alpha x}L_{\alpha y}$ which necessarily has many negative components, so gradient ascent will bring the network to a configuration with many negative geometric stress components. 

\begin{figure}
\centering
\includegraphics[scale= 1]{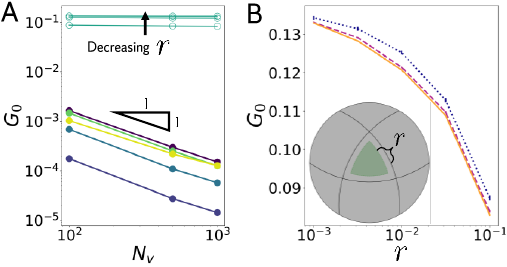}
\caption{\textbf{Constrained optimization of the shear modulus $G_0$.} A) Size scaling of the optimal $G_0$ for all groups of critical configurations. Configurations that were not optimized for shear modulus (filled circles) have the expected finite-size scaling for isotropic systems $G_0 \sim 1/N$, while configurations that were optimized for $G_0$ (open circles) do not scale with system size. B) Optimal values of $G_0$ for various thresholds $r$ and for system sizes $N_v=100$ (dotted), $500$ (dashed) and $1000$ (solid). As an example, the inset illustrates the region of geometric stresses of the three bar linkage that satisfy $\sigma^h > r$ (colored green).}
\label{ShearModulusScaling}
\end{figure}

We can see that the edges begin growing to an extent that would be difficult to realize in an experiment, with system-spanning edges and edge overlaps, at the same point in the optimization that the smallest tensions become negative. This suggests that to study configurations that are possible to realize in an experiment, we might add additional constraints on the geometric stress to keep the edges under tension rather than compression. Therefore, we add an additional term to the objective function that penalizes components of $\sigma^h_\alpha$ that are less than a given threshold value $r>0$,
\begin{align}
    \mathcal{O}(\sigma^h, r) = G_0(\sigma^h) - \frac{1}{N_b}\sum_\alpha \left(\frac{\sigma^h_\alpha}{r}\right)^{-10}, \label{ShearMod+Constraint}
\end{align}
where the normalization of the geometric stress is fixed as $\sum_\alpha (\sigma^h_\alpha)^2 = N_b$. An illustration of this constraint in geometric stress space is shown in the inset to Fig.~\ref{ShearModulusScaling}B, with the allowed region highlighted in teal.
In Fig. \ref{Networks} we show networks that result from maximizing Eq. (\ref{ShearMod+Constraint}) with various values of the threshold $r$. The open teal circles in Fig.~\ref{ShearModulusScaling}A demonstrates that this constrained optimization does generate systems with finite $G_0$, independent of system size $N_v$, in contrast to other critical networks (closed circles in Fig.~\ref{ShearModulusScaling}A) which scale as $1/N_v$. Lower values of this threshold result in the network becoming increasingly aligned along the shear direction.

\subsubsection{Structural Objective Functions}
Suppose we wanted to make a second-order rigid network that can be assembled by combining many identical components. In this case, we would like a critical configuration where the edges lengths are as similar as possible. We can look for such configurations by minimizing the relative fluctuations of the lengths:
\begin{align}
    V_L &= \frac{\boldsymbol\langle (L - \langle L \rangle)^2 \boldsymbol\rangle}{\langle L\rangle^2} = \frac{\left\langle L^2 \right\rangle}{\langle L\rangle^2} - 1,\label{LengthVariance}
\end{align}
where $\langle \rangle$ indicates an average over the edges of the network. Similarly we could minimize the fluctuations of the putative edge tensions $\tau_\alpha = \sigma^h_\alpha L_\alpha$, which are the edge lengths of the dual force network.
\begin{align}
    V_\tau &= \frac{\boldsymbol\langle (\tau - \langle \tau \rangle)^2 \boldsymbol\rangle}{\langle \tau\rangle^2} = \frac{\left\langle \tau^2 \right\rangle}{\left\langle \tau \right\rangle^2} - 1 = \frac{\langle (\sigma^h L)^2 \rangle}{\left\langle \sigma^h L \right\rangle^2} - 1.\label{TensionVariance}
\end{align}
This is potentially useful for design, as if the edges of the network can only support a finite amount of tension before breaking, a network with smaller fluctuations in tensions would be more resistant to fracture.

We find that each network structure has a single unique critical configuration that minimizes $V_L$, shown in Fig~\ref{Networks}A. In contrast, there are many different configurations that minimize $V_\tau$, and several examples with the same $V_\tau$ are shown in Fig~\ref{Networks}B,C. This is because $V_\tau$ depends only on the dual force network, and two different configurations of the physical network possess the same force network when they are related by an isogonal mode which changes the lengths of the edges without changing the angles between them \cite{noll_active_2017, brauns_geometric_2024}.

\subsubsection{Random Critical Configurations}

We also study two populations of critical configurations from different random samples of critical configurations. First, we randomly assign rest lengths to the edges of the network and uniformly decrease them (equivalent to performing a volumetric expansion) until the system hits the critical manifold and becomes rigid. This is the standard method used to study strain stiffening in spring and fiber network model. Examples are shown in Fig.~\ref{Networks}J,K. We also randomly assign the component of geometric stress associated with each edge and use the geometric stress parameterization to obtain the corresponding critical configuration. Details of these sampling procedures can be found in the Supplemental Materials. Examples are shown in Fig.~\ref{Networks}L,M.

\begin{figure}
\centering
\includegraphics[scale= 0.32]{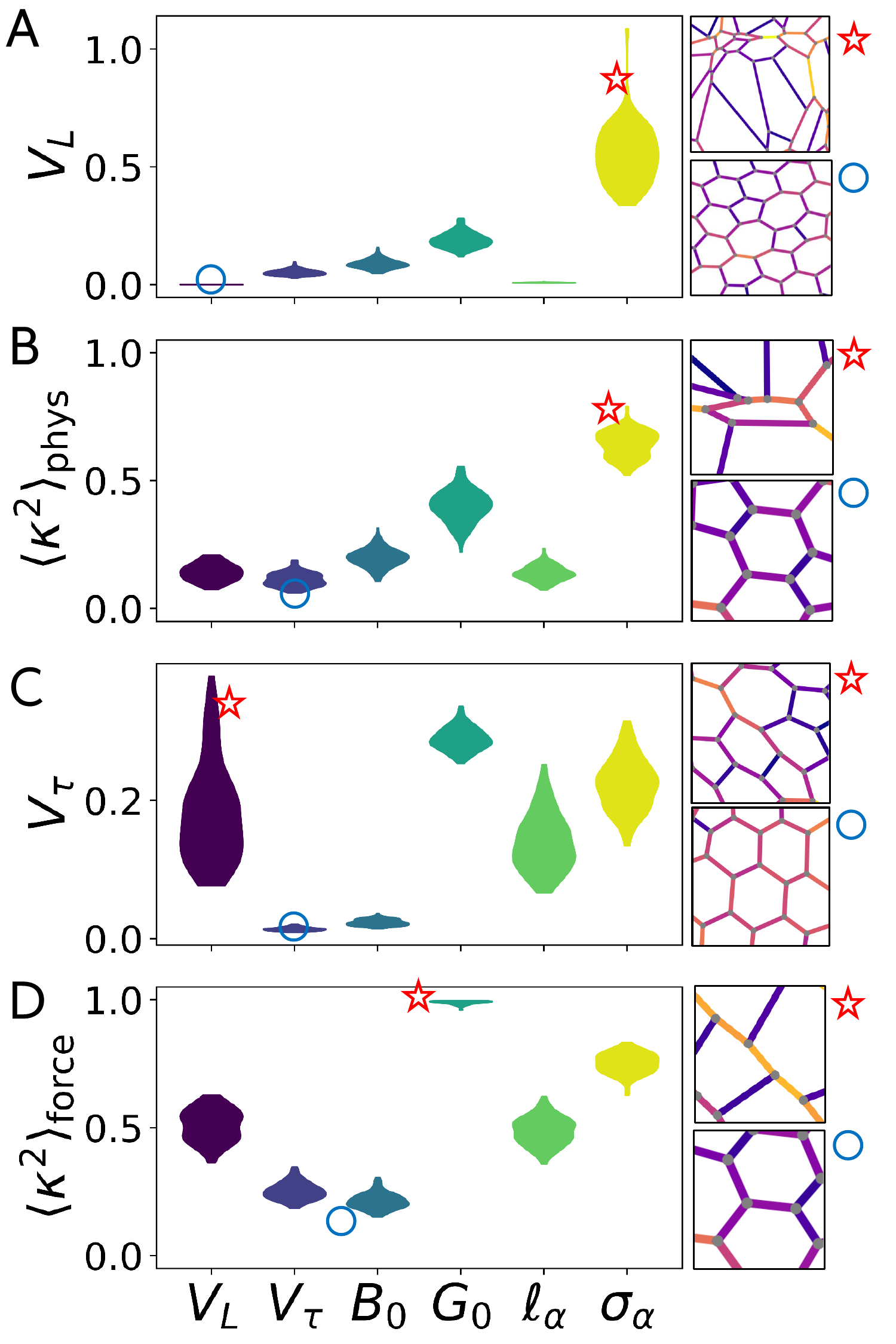}
\caption{Structural Metrics of Different Populations of Critical Configurations: The horizontal axis labels $V_L$, $V_\tau$, $B_0$, and $G_0$ refer to the objective function being optimized, while $\ell_\alpha$ and $\sigma^h_\alpha$ refer to the populations derived by randomly assigning rest lengths and geometric stresses respectively. The images of networks next to the violin plots show examples of network structures that correspond with high and low values of each of the structural metrics.}
\label{StructuralMetrics}
\end{figure}

\subsection{Analyzing Ensemble Network Properties}
We seek to understand ensemble-averaged properties for these different types of configurations. For each instance of 100 unique network structures described by incidence matrix $g_{\alpha i}$ with $N_v=1000$ vertices, we generate six populations of critical configurations: minimized $V_L$, minimized $V_\tau$, maximized $B_0$, maximized $G_0$, random rest lengths, and random geometric stresses. \tah{Note that for the purpose of comparing the configurations that were optimized for $G_0$ with the other networks, we use $r=0.1$ for the threshold on the geometric stress, an example of which is shown in Fig. \ref{Networks}G. }

We then compare these populations in order to investigate the full range of behaviors and properties that these second-order rigid networks can achieve, and look for emergent properties of the optimized networks that could eventually be used to generate such structures using local rules instead of global optimization. 


\subsubsection{Local Structure}
First, we compare the local structure of the configurations in both physical space and in the dual force space. In Figure \ref{StructuralMetrics}, we show for each population of critical configurations violin plots of the distributions of several structural parameters. First the objective function $V_L$, which quantifies the fluctuations in the edge lengths. As this quantity does not measure the orientations of the edges, we also plot the average shape anisotropy of the faces of the networks $\langle\kappa_{\text{phys}}\rangle$. The shape anisotropy of a face with vertices indexed by $i=1,..,n$ with coordinates $r^i_\mu$ can be written in terms of the eigenvalues $\lambda_1$, $\lambda_2$ of the gyration tensor $S_{\mu\nu} = 1/n \sum_i r^i_\mu r^i_\nu$:
\begin{align}
    \kappa_{\text{phys}} = 2\frac{\lambda_1^4 + \lambda_2^4}{(\lambda_1^2 + \lambda_2^2)^2} -1  \label{ShapeAnisotropy}
\end{align}
This quantity is 0 when the face is a regular polygon, and close to 1 if the face is elongated along a particular direction \cite{sarkar_shear-induced_2016}. Note that the average $\langle\kappa_{\text{phys}}\rangle$ is taken over all faces of a single member of a population, and the violin plots in \ref{StructuralMetrics} show the distribution of this quantity over each ensemble. We then plot the equivalents of the length fluctuations and shape anisotropy for the dual force networks $V_\tau$ and $\langle\kappa_{\text{force}}\rangle$. 

We see that for all metrics, networks with optimized $V_L$ are very similar those with random rest lengths, which makes sense as the random rest lengths are all chosen from the same Gaussian distribution. \tah{Our goal when optimizing $V_L$ was to find a critical configuration with equal lengths, and thus a variance $V_L = 0$. With our chosen network structure, the optimization procedure is able to find critical configurations with very small variances in edge lengths on the order of $V_L \lesssim 10^{-7}$.} \mlm{In other words, for \textbf{any} given incidence matrix in our ensemble, it is generically possible for us to identify a single spring length $L^*$ so that if the network were assembled of units of that length (with \textbf{any} type of nonlinear spring response), the material would be precisely at the rigidity transition.  Note that in this work we study a limited ensemble of incidence matrices that are three-fold random regular coordinated, with initial structure generated by a random sequential addition protocol that further regularizes the network topology, and future work could expand this ensemble.}

Networks with optimized $B_0$ are also largely similar to those with optimized $V_\tau$, as both have small variations in edge tensions, leading to regular force networks and hence small values of $\langle\kappa_{\text{force}}\rangle$. \tah{However, unlike the configurations that were able to achieve negligible values of $V_L$, the networks with optimal $V_\tau$ still have a non-zero variance in edge tensions $V_\tau \approx 10^{-2}$. This is because any network with equal edge tensions must have all adjacent edges meet at $120 \degree$ angles in order to maintain force balance, and such an arrangement is only geometrically possible in a regular hexagonal lattice and not in a generic 3-regular graph.} The other populations of configurations have edges that are more aligned, which causes a concentration of tension. Configurations with optimized $G_0$ (with a geometric stress tolerance of $r=0.1$) are highly aligned, and $\langle\kappa_{\text{force}}\rangle$ is close to its maximum value of 1 as the triangles that make up the dual force network are nearly one-dimensional. This avoids the near-complete alignment of edges that occurs as this threshold is decreased, as can be seen in \ref{Networks}H-I.

\subsubsection{Mechanical Response}
We also characterize the mechanical behavior of the different populations of critical configurations as they are strained away from the critical manifold. Using the same ensembles as in Fig. \ref{StructuralMetrics}, we start each configuration on the critical manifold then subject the networks to a volumetric expansion by increasing the volume as
\begin{align}
    V(\gamma) = V(0)(1+\gamma) = N_v(1+\gamma), \label{RestLengthVolumetric}
\end{align}
in strain steps of $\delta\gamma = 0.001$. At each strain step we minimize the elastic energy in Eq. (\ref{HookeanEnergy}) as a function of the vertex coordinates $x_{i\mu}$ using the FIRE algorithm until the largest component of the unbalanced forces $\partial E/\partial x_{i\mu}$ has a magnitude less than $10^{-15}$. Using linear response, we then calculate the bulk and shear modulus at each strain step.

\begin{figure}
\centering
\includegraphics[scale= 0.32]{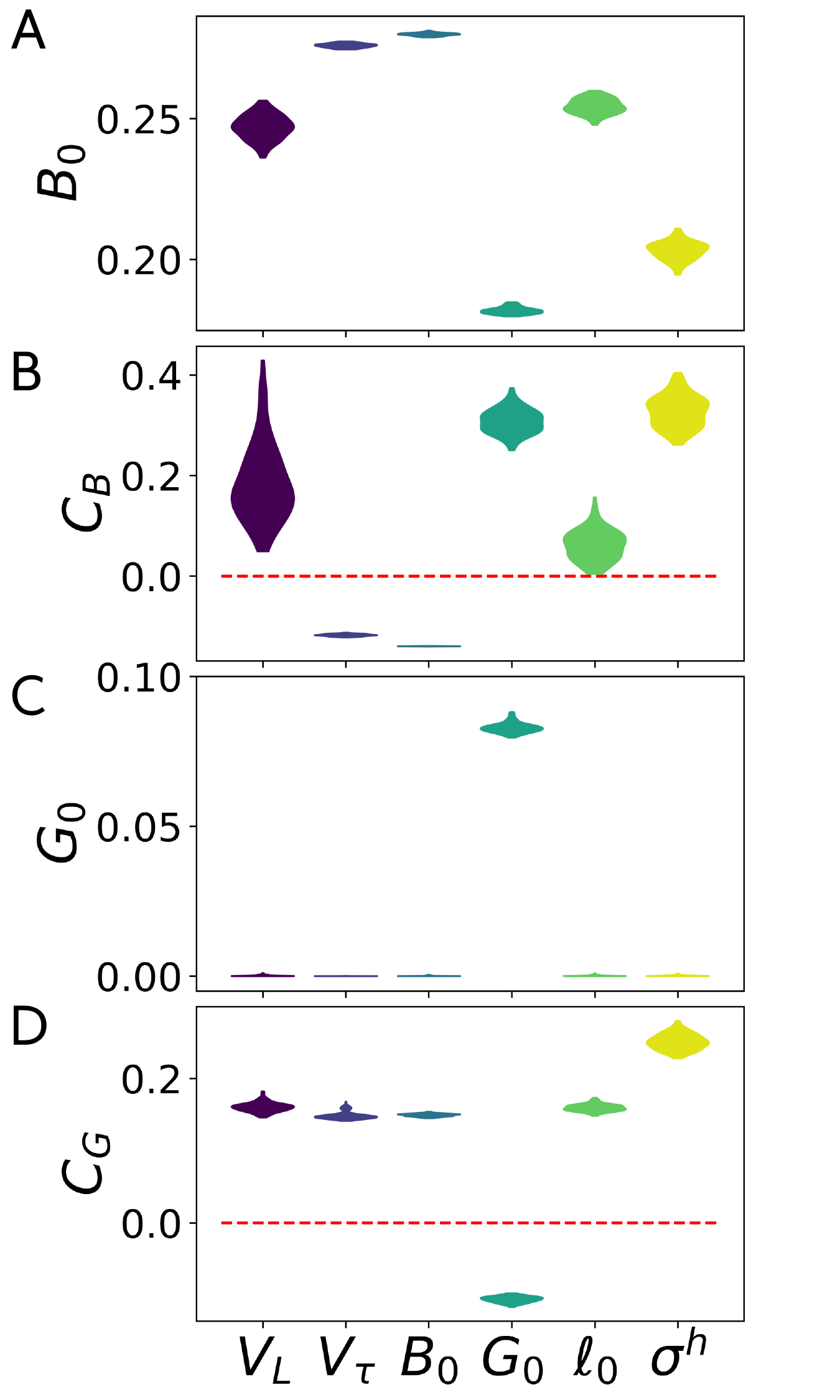}
\caption{\textbf{Mechanical Response Metrics of Different Populations of Critical Configurations.} Distributions of A) the bulk modulus $B_0$ at the transition point, B) the bulk modulus strain scaling coefficient $C_B$, C) the shear modulus $G_0$ at the transition point, and D) the shear modulus strain scaling coefficient $C_G$ for the same six ensembles of critical configurations as in Fig. \ref{StructuralMetrics}. The red lines in B,D) are drawn at $C_{B,G} = 0$ to visually separate populations whose elastic moduli increase or decrease with volumetric strain.}
\label{ResponseMetrics}
\end{figure}

\tah{In Fig. \ref{ResponseMetrics}A \& C we plot the distributions of the bulk modulus $B_0$ and shear modulus $G_0$ at the transition point. As expected, the networks that were optimized for $B_0$ do in fact have the largest bulk moduli, followed closely by the networks optimized for $V_\tau$. This maximum value of the bulk modulus is about $10\%$ greater than that of networks that were assigned random rest lengths or were optimized for $V_L$, and about $50\%$ higher that of configurations optimized for $G_0$ or derived from a random geometric stress. Networks optimized for shear modulus have significantly larger $G_0$ than the other populations of approximately isotropic configurations, which have negligible shear moduli due to finite size effects that vanish in the thermodynamic limit as $G_0 \sim N^{-1}$ as shown in Fig. \ref{ShearModulusScaling}. }

\tah{ We also investigate how the elastic moduli change as the networks are strained past the rigidity transition. Close to the critical point, the moduli should scale linearly with volumetric strain \cite{lee_stiffening_2022}
\begin{align}
    B(\gamma) &= B_0 + C_B \gamma \nonumber,\\
    G(\gamma) &= G_0 + C_G \gamma. \label{ModulusStrainScaling}
\end{align} 
Fig. \ref{ResponseMetrics}B \& D show the distributions of the coefficients $C_B$ and $C_G$, which were obtained from a least-squares fit of the moduli measured near the transition. We observe that populations with larger moduli have smaller scaling coefficients. Networks with optimized $B_0$ or $V_\tau$ have the largest bulk moduli at the transition, but this decreases with additional strain (i.e $C_B < 0$), whereas the bulk modulus of the other types of configurations increases with strain $(C_B > 0)$. Similarly, networks optimized for $G_0$ have their shear moduli decrease with strain, and all other groups' shear moduli increase past the transition point. }

\section{Summary and Discussion}
In this paper we have described a method to parameterize the space of all critical configurations of any central force network. We showed that we can use gradient descent to search the space of geometric stresses to find critical configurations with particular structural properties (equal edge lengths or tensions) or elastic responses (maximum bulk or shear modulus). \mlm{For the shear modulus, we found that global optimization pushes the system towards extreme states that are not experimentally realizable, though it is straightforward to constrain the minimization to reasonable states that exhibit finite $G_0$ even in the large-$N$ limit, in contrast to random disordered networks where $G_0$ vanishes.}

The geometric stress parameterization allows us to enumerate all possible rigid configurations of a given network structure. In this paper we have focused on 3-coordinated random regular networks for simplicity, but future work should investigate how changes to the topology of a network affect the form of the critical manifold. This could be useful to study how underconstrained networks fracture when edges with high tension are broken~\cite{driscoll2016role}, or how biopolymer networks are remodelled by fiber-cleaving enzymes which preferentially sever low-tension fibers~\cite{saini2020tension}.

\mlm{One result that may be particularly useful for materials design is our discovery that we can design materials perched precisely at the rigidity transition that are composed of units of identical length (e.g. we can optimize so that $V_L \sim 0$ for all network topologies in our limited ensemble).  This suggests that it may be possible to design self-assembled materials where the network topology is programmed, e.g. by small DNA tags, to link together "bars" that are all a standard length. Because the network would sit exactly at the edge of a rigidity transition, the resulting material could be stiffened by orders of magnitude via very small changes to tuning parameters, and this macroscopic, emergent property would be robust to small differences in the mechanical properties of the individual bars. Moreover, with small extensions to the work presented above, it should also be possible to program materials that strain stiffen or become floppy after the material experiences a precisely controlled amount of strain.}

In overconstrained networks, the force transmission properties are relatively divorced from the precise structural details, as any system will have many different ways to distribute stress through the system. However, the geometric stress parameterization shows that in underconstrained networks the structure and force transmission are intimately connected, both arising from the underlying geometry of the critical manifold. This has implications for the possible states that underconstrained networks can access. For example, several researchers have looked at emergent alignment of stress in underconstrained networks that are strained to the critical point~\cite{abhilash2014remodeling,taufalele2019fiber, wang2014long}. In future work we plan to study how the geometry of the critical manifold makes aligned structures highly probable.

Underconstrained networks also differ from overconstrained networks in that their vibrational spectrum is separated into two distinct groups. There are ``longitudinal" modes that involve first-order deformations to the members of the network and whose behavior are thus determined in part by material properties, and there are ``transverse" modes (the linear zero modes) which are governed by geometry alone. It has been shown that in ordered networks with sufficient symmetries, these populations of modes can be completely decoupled \cite{rocks_integrating_2024}. The geometric stress parameterization gives a framework for searching for such states in a disordered network, which would then have interesting vibrational behaviour.

\tah{Mechanical networks have been trained to exhibit auxetic behavior \cite{rocks_designing_2017, reid_auxetic_2018} as well as specific allostery-inspired responses \cite{rocks_designing_2017}.} However, as these learning methods rely on the rigidity of the network, most studies of physical learning algorithms in mechanical networks have focused on overconstrained networks. \mlm{The geometric stress parameterization should be very helpful in implementing learning rules and global gradient descent that optimize desired objective functions in underconstrained networks, and may therefore accelerate research on physical learning in such systems.}

\section{Acknowledgements}
MLM and TH were supported by Simons Foundation \#454947, NSF-DMR-1951921 and SU’s Orange Grid research computing cluster. CDS was supported by DMR-2217543.
\bibliographystyle{unsrt}
\bibliography{MyLibrary}




\section{Supplemental Materials}

\subsection{Generating Random Underconstrained Networks}
In the main text, we compared the emergent properties of four different populations of critical configurations of 2 dimensional periodic central-force networks. These networks were initially generated by making a Voronoi tessellation using the algorithm in the Qhull library implemented in SciPy. The sets of points used to produce the Voronoi diagrams were generated by a random sequential addition procedure. The coordinates of the points are chosen randomly from a unit square one at a time, and each successive point is accepted only if it further than a distance $r$ away from all previously accepted points. Varying the threshold distance $r$ changes the total number of points that can be placed. 

We use this procedure to produce networks that experience full rigidity percolation under strain when all edges are given similar rest lengths, in order to compare them with the fully rigid critical configurations obtained from the geometric stress parameterization. If the seed points for the Voronoi tessellation are chosen purely randomly, the resulting networks have large fluctuations in the density of vertices, and regions with higher density then become rigid at a larger critical strain than low-density regions when all edges have similar rest lengths.

Once the Voronoi tessellation has been generated, giving us a network with $N_v$ vertices and $N_b = 3N_v/2$ edges, we describe the underlying graph structure with the incidence matrix. After choosing an arbitrary orientation for each edge, the incidence matrix $g_{\alpha i}$ is given by
\begin{align}
    g_{\alpha i} = \begin{cases}
        +1 \quad\text{if vertex } i \;\text{is the head of edge } \alpha \\
        -1 \quad\text{if vertex } i \;\text{is the tail of edge } \alpha \\
        \;\;0 \;\quad\text{else}
    \end{cases}
    \label{incidenceMatrix}
\end{align}

\begin{figure}
\centering
\includegraphics[scale=1]{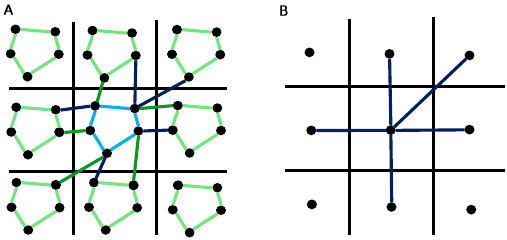}
\caption{A) A small periodic network. Light blue edges connect two vertices within the principle domain, with their periodic copies in light green. Dark blue edges connect vertices in different periodic copies, crossing the periodic boundary (periodic copies shown in dark green). B) The same network with all vertices placed at the same point, demonstrating the boundary term $b_{\alpha\mu}$.}
\label{BoundaryTerm}
\end{figure}

If the $\mu = x,y$ coordinate of vertex $i=1,..,N_v$ is $x_{i\mu}$, we can write the $\mu$ component of the vector along edge $\alpha$ as
\begin{align}
    L_{\alpha\mu} = \sum_i g_{\alpha i}x_{i\mu} + b_{\alpha\mu}, \label{BondVectors2}
\end{align}
where $b_{\alpha\mu}$ encodes the periodic boundaries as shown in Figure \ref{BoundaryTerm}. The light blue interior edges have $b_{\alpha\mu}=0$, while the dark blue edges that cross the periodic boundary have non-zero $b_{\alpha\mu}$. For example, the edge connecting the principle domain to the 
periodic copy directly to the right would have $b_{\alpha\mu}=[S, 0]$ where $S$ is the linear dimension of the periodic box, which we take to be $S=\sqrt{N_b}$ so that the area of the box scales with the number of vertices and edges. If we want to apply an affine deformation $A_{\mu\nu}$ to the network, we can simply apply it to the boundary term: $b_{\alpha\mu} \rightarrow A_{\mu\nu}b_{\alpha\nu}$.

\subsection{Analytic Expressions}

\subsubsection{Deriving Rigidity, Geometric Response, and Prestress Matrices}
We can now derive the expressions for the geometric response matrix and the prestress matrix by taking derivatives of the quantity $h_\alpha = L_{\alpha}^2/2$.
\begin{align}
    R^h_{\alpha i\mu} &= \frac{\partial h_\alpha}{\partial x_{i\mu}} = \frac{1}{2}\sum_{\nu}\frac{\partial L_{\alpha\nu}^2}{\partial x_{i\mu}} = \sum_{\nu}\frac{\partial L_{\alpha\nu}}{\partial x_{i\mu}}L_{\alpha\nu} \nonumber\\
    &= g_{\alpha i}\delta_{\mu\nu}L_{\alpha\nu} = g_{\alpha i}L_{\alpha\mu},
\end{align}\label{RigidityMatrix2}\begin{align}
    P_{i\mu j\nu} &= \sum_\alpha \sigma^h_\alpha\frac{\partial^2 h_\alpha}{\partial x_{i\mu}\partial x_{j\nu}} = \sum_\alpha \sigma^h_\alpha \frac{\partial}{\partial x_{j\nu}}\left(g_{\alpha i}L_{\alpha\mu}\right) \nonumber \\ 
    &= \sum_\alpha \sigma^h_\alpha g_{\alpha i}g_{\alpha j} \delta_{\mu\nu} = P_{ij}\delta_{\mu\nu}.
\end{align}
\label{PrestressMatrix2}

As an illustrative example, let us derive these matrices for the Hookean constraint $f_\alpha = L_\alpha - \ell_\alpha$:
\begin{align}
    \tilde{R}_{\alpha i\mu} &= \frac{\partial \tilde{f}_\alpha}{\partial x_{i\mu}} = \frac{\partial L_{\alpha}}{\partial x_{i\mu}} = \frac{1}{2L_\alpha}\frac{\partial L_{\alpha}^2}{\partial x_{i\mu}}L_{\alpha\nu} = g_{\alpha i}\frac{L_{\alpha\mu}}{L_\alpha}, \label{HookeanRigidityMatrix}
\end{align}
\begin{align}
    \tilde{P}_{i\mu j\nu} &= \sum_\alpha \tau_\alpha\frac{\partial^2 \tilde{f}_\alpha}{\partial x_{i\mu}\partial x_{j\nu}} = \sum_\alpha \tau_\alpha \frac{\partial}{\partial x_{j\nu}}\left(g_{\alpha i}\frac{L_{\alpha\mu}}{L_\alpha}\right) \nonumber\\
    &= \sum_\alpha \tau_\alpha g_{\alpha i}\left(\frac{1}{L_\alpha}\frac{\partial L_{\alpha\mu}}{\partial x_{j\nu}} - \frac{L_{\alpha\mu}}{L_\alpha^2}\frac{\partial L_{\alpha}}{\partial x_{j\nu}} \right) \nonumber\\
    &= \sum_\alpha \frac{\tau_\alpha}{L_\alpha} g_{\alpha i}g_{\alpha j}\delta_{\mu\nu} - \frac{\tau_\alpha}{L_\alpha^3}L_{\alpha\mu}g_{\alpha i}g_{\alpha j}L_{\alpha\nu} \nonumber\\
    &= P_{i\mu j\nu} - \sum_\alpha \tilde{R}_{\alpha i\mu}\tau_\alpha \tilde{R}_{\alpha j\nu}.
    \label{HookeanPrestressMatrix}
\end{align}

Since $R$ and $\tilde{R}$ are identical up to a rescaling of the rows by $L_\alpha$, they have the same right nullspace, so the linear zero modes are exactly the same. The prestress matrix derived from the Hookean constraint $\tilde{P}$ is the same as the prestress matrix of the squared lengths up to an additional term that depends on the rigidity matrix. However, because the rigidity of the network is determined by the projection of the prestress matrix onto the subspace of linear zero modes (which are by definition in the nullspace of $R$), this extra term adds no new information about the rigidity of the network.

\subsubsection{Geometry of the Critical Manifold}
In order to calculate total derivatives of an objective function, we need to write the derivatives of the geometric stress parameterization. Let us write the bond vector $L_{\beta\mu}$ as a function of $\sigma^h_\alpha$ using the geometric stress parameterization:
\begin{align}
    L_{\beta\mu} &= \sum_i g_{\beta i}x_{i\mu} + b_{\beta\mu} \\
    &= \sum_i g_{\beta i}\left[\sum_{j\gamma} -\left(P^{-1}\right)_{ij}g_{\gamma j}\sigma^h_\gamma b_{\gamma\mu} \right] + b_{\beta\mu} \\
    &= -\sum_{ij\gamma}g_{\beta i} \left(P^{-1}\right)_{ij} g_{\gamma j}\sigma^h_\gamma b_{\gamma\mu} + b_{\beta\mu},
    \label{BondVectorFDM}
\end{align}
where by $P$ we mean the non-zero $N_v\times N_v$ block $P_{ij} = \sum_\alpha \sigma_\alpha^h g_{\alpha i}g_{\alpha j}$ of the full $2N_v\times 2N_v$ prestress matrix $\bm P_{i\mu j\nu} = P_{ij}\delta_{\mu\nu}$. Taking a derivative with respect to $\sigma^h_\alpha$ gives
\begin{widetext}
\begin{align}
    \frac{\partial L_{\beta\mu}}{\partial \sigma^h_\alpha} &= \frac{\partial}{\partial \sigma^h_\alpha} \left[-\sum_{ij\gamma}g_{\beta i} \left(P^{-1}\right)_{ij} g_{\gamma j}\sigma^h_\gamma b_{\gamma\mu} + b_{\beta\mu}\right] \\
    &= \sum_{ijkl\gamma}g_{\beta i} \left(P^{-1}\right)_{ij} \frac{\partial P_{jk}}{\partial \sigma^h_\alpha} \left(P^{-1}\right)_{kl} g_{\gamma l}\sigma^h_\gamma b_{\gamma\mu} - \sum_{ij}g_{\beta i} \left(P^{-1}\right)_{ij} g_{\alpha j} b_{\alpha\mu} \\
    &= \left[\sum_{ij}g_{\beta i} \left(P^{-1}\right)_{ij} g_{\alpha j}\right] \left[\sum_{kl\gamma}g_{\alpha k} \left(P^{-1}\right)_{kl} g_{\gamma l}\sigma^h_\gamma b_{\gamma\mu}\right] - \left[\sum_{ij}g_{\beta i} \left(P^{-1}\right)_{ij} g_{\alpha j}\right] b_{\alpha\mu} \\
    &= -\left[\sum_{ij}g_{\beta i} \left(P^{-1}\right)_{ij} g_{\alpha j}\right]\left[-\sum_{kl\gamma}g_{\alpha k} \left(P^{-1}\right)_{kl} g_{\gamma l}\sigma^h_\gamma b_{\gamma\mu} + b_{\alpha\mu}\right] = -\left[\sum_{ij}g_{\beta i} \left(P^{-1}\right)_{ij} g_{\alpha j}\right]L_{\alpha\mu},
    \label{BondVectorDerivativeFDM2}
\end{align}
\end{widetext}

\newpage

where we have used the standard matrix identity
\begin{align}
    \frac{\partial}{\partial x} P^{-1} = -P^{-1} \frac{\partial P}{\partial x} P^{-1}.
\end{align}
Now we can write the derivative of the squared lengths $h_\beta = L_\beta^2/2$:
\begin{align}
    &\frac{\partial h_\beta}{\partial \sigma^h_\alpha} = \sum_\mu L_{\beta\mu}\frac{\partial L_{\beta\mu}}{\partial \sigma^h_\alpha}
    \\ &= -\sum_\mu L_{\beta\mu}\left[\sum_{ij\mu}g_{\beta i} \left(P^{-1}\right)_{ij} g_{\alpha j}\right]L_{\alpha\mu}.
    \label{BondVectorDerivativeFDM}
\end{align}
Since the geometric response matrix is $R^h_{\alpha i\mu} = g_{\alpha i}L_{\alpha\mu}$, we can further write this as the matrix product
\begin{align}
\frac{\partial h_\beta}{\partial \sigma^h_\alpha} = -\left(R^h \bm P^{-1}{R^h}^T\right)_{\alpha\beta}.\label{tangents}
\end{align}
Then $\sum_\beta \sigma^h_\beta (\partial h_\beta /\partial \sigma^h_\alpha)$ must vanish because $\sigma^h_\beta$ is guaranteed to be in the left nullspace of $R^h$. As discussed in the main text, the columns of this matrix span the tangent space of the critical manifold $h_\beta(\bm\sigma^h)$ at the current configuration, so this implies that the geometric stress is normal to the critical manifold.

We can also calculate the second fundamental form of the critical manifold, which describes the local curvature and is given by

\begin{align}
    \mathrm{I\!I}_{\beta\gamma} = \sum_\alpha \frac{\partial^2 h_\alpha}{\partial s_\beta \partial s_\gamma}\frac{\sigma^h_\alpha}{|\bm\sigma^h|} = \frac{1}{|\bm\sigma^h|} (V^T B V)_{\beta\gamma},
    \label{2FF2}
\end{align}
where $s_\beta$ are the $N_b-1$ stereographic coordinates as described in the main text and $\sigma^h_\alpha/|\bm\sigma^h|$ is the local unit normal vector for the manifold. Defining $V_{\alpha\beta} := \partial \sigma^h_\alpha/\partial s_\beta$ we have
\begin{align}
    \mathrm{I\!I}_{\beta\gamma} &= \frac{1}{|\bm\sigma^h|}\sum_{\delta \epsilon} \frac{\partial \sigma^h_\delta}{\partial s_\beta} \left( \sum_\alpha \sigma^h_\alpha \frac{\partial^2 h_\alpha}{\partial \sigma^h_\delta \partial \sigma^h_\epsilon} \right) \frac{\partial \sigma^h_\epsilon}{\partial s_\beta} \\
    &= \frac{1}{|\bm\sigma^h|} (V^T B V)_{\beta\gamma},
\end{align}
where $B_{\beta\gamma} = \sum_\alpha \sigma^h_\alpha (\partial^2 h_\alpha / \partial \sigma^h_\beta \partial \sigma^h_\gamma)$ satisfies the interesting equality
\begin{widetext}
\begin{align}
    B_{\beta\gamma} &= \sum_\alpha \sigma^h_\alpha \frac{\partial^2 h_\alpha}{\partial \sigma^h_\beta \partial \sigma^h_\gamma} = \sum_\alpha \sigma^h_\alpha \frac{\partial }{\partial \sigma^h_\beta} \left( -R^h P^{-1}{R^h}^T\right)_{\alpha\gamma} \\
    &= -\sum_\alpha \sigma^h_\alpha \left[ \left(\frac{\partial R^h}{\partial \sigma^h_\beta}\right)P^{-1} {R^h}^T + R^h \left(\frac{\partial P^{-1}}{\partial \sigma^h_\beta}\right){R^h}^T + R^h P^{-1}\left(\frac{\partial {R^h}^T}{\partial \sigma^h_\beta}\right) \right]_{\alpha\gamma} \\
    &= -\sum_{ij\mu}\sum_\alpha \sigma^h_\alpha \left[\frac{\partial }{\partial \sigma^h_\beta}\left( g_{\alpha i}L_{\alpha\mu} \right)\right] \left(P^{-1}\right)_{ij} g_{\gamma j}L_{\gamma\mu} = \sum_{ij\mu}\sum_\alpha \sigma^h_\alpha g_{\alpha i} \left[\frac{\partial L_{\alpha\mu}}{\partial \sigma^h_\beta}\right] \left(P^{-1}\right)_{ij} g_{\gamma j}L_{\gamma\mu}\\
    &= \sum_{ij\mu}\sum_\alpha \sigma^h_\alpha g_{\alpha i} \left[\sum_{kl}g_{\alpha k} \left(P^{-1}\right)_{kl} g_{\beta l}L_{\beta\mu}\right] \left(P^{-1}\right)_{ij} g_{\gamma j}L_{\gamma\mu} \\
    &= \sum_{lj\mu} (g_{\beta l}L_{\beta\mu}) \left[ \left(P^{-1}\right)_{kl}\left(\sum_\alpha \sigma^h_\alpha \;g_{\alpha i}g_{\alpha k}\right)\left(P^{-1}\right)_{ij} \right] (g_{\gamma j}L_{\gamma\mu})\\
    &= \sum_{lj\mu} R^h_{\beta l\mu}\left(P^{-1}P P^{-1}\right)_{lj}R^h_{\gamma j\mu} = \left(R^h \bm P^{-1} {R^h}^T\right)_{\beta\gamma} = \frac{\partial h_\beta}{\partial \sigma^h_\gamma},
\end{align}
\end{widetext}
using Eqs. \ref{tangents} \& \ref{BondVectorDerivativeFDM} and the fact that by definition the geometric stress $\sigma_\alpha^h$ is in the nullspace of the geometric response matrix $R^h$. So for the critical manifold the matrix $B$ determines both the local curvature through the second fundamental form as well as the gradients $\partial h_\beta/\partial \sigma_\gamma^h$, which give the local tangent space and hence the metric.

\subsubsection{Elastic Moduli at the Rigidity Transition}
We want to write the elastic moduli of a network as a simple function of the network configuration $L_{\alpha\mu}$ and the geometric stress $\sigma^h_\alpha$ in order to take derivatives and perform gradient descent. Consider a global affine deformation $I + \gamma A_{\mu\nu}$, which changes the boundary term and the bond vectors as
\begin{align}
\frac{\partial b_{\alpha\mu}}{\partial \gamma} &= \sum_\nu A_{\mu\nu}b_{\alpha\nu} \\
\frac{\partial L_{\alpha\mu}}{\partial \gamma} &= \sum_\nu A_{\mu\nu}L_{\alpha\nu},
\end{align}
which results in a change to the squared lengths $h_\alpha = \sum_\mu L_{\alpha\mu}^2/2$
\begin{align}
\dot{h} := \frac{\partial h_{\alpha}}{\partial \gamma} = \sum_\mu L_{\alpha\mu} \frac{\partial L_{\alpha\mu}}{\partial \gamma} = \sum_{\mu\nu} L_{\alpha\mu}A_{\mu\nu}L_{\alpha\nu}.
\end{align}
The elastic modulus associated with this deformation can be calculated as
\begin{align}
\mu_A = \frac{1}{V}\frac{\text{d}^2 E}{\text{d} \gamma^2} = \frac{1}{V}\left(\frac{\partial^2 E}{\partial \gamma^2} - \bm\Xi^T H^{-1} \bm\Xi \right),
\end{align}
where $V$ is the system volume (or area for 2D systems). The first term contains the contributions from the affine motions of the system, while the second term accounts for the non-affine relaxation that occurs in order to achieve force balance. $H = \partial^2 E/\partial x_i \partial x_j$ is the Hessian matrix and $\Xi_i = \partial^2 E/\partial x_i \partial \gamma$ contains the unbalanced forces that appear as a result of the affine deformation. Note that $H^{-1}$ should be understood as the pseudo-inverse of $H$, which projects out any zero modes. Let us calculate these quantities at the second order rigidity transition, where a self-stress exists but the prestress is zero $(\partial E/\partial h_\alpha = 0)$. First the affine term:
\begin{align}
\frac{\partial^2 E}{\partial^2 \gamma} &= \frac{\partial}{\partial \gamma}\left(\sum_\alpha \frac{\partial E}{\partial h_\alpha} \frac{\partial h_\alpha}{\partial \gamma} \right) \\ \nonumber
&= \sum_{\alpha\beta} \dot{h}_\alpha \frac{\partial^2 E}{\partial h_\alpha \partial h_\beta}\dot{h}_\beta + \sum_\alpha \frac{\partial E}{\partial h_\alpha} \frac{\partial^2 h_\alpha}{\partial \gamma^2} \\ \nonumber
&= \dot{\bm{h}}^T K \dot{\bm{h}},
\end{align}
where $\dot{h}_\alpha = \partial h_\alpha/\partial \gamma$ gives the change in squared lengths if the network due to the chosen affine deformation and $K_{\alpha\beta} = \partial^2 E/\partial h_\alpha \partial h_\beta$ is the matrix of modified stiffnesses. For a network of Hookean springs with energy $E = \sum_\alpha k_\alpha(L_\alpha - \ell_\alpha)^2/2$ at zero prestress, this is a positive-definite diagonal matrix: $K_{\alpha\beta} = k_\alpha/L_\alpha^2 \;\delta_{\alpha\beta}$. Next, at the transition the Hessian is given by just the Gram term
\begin{align}
H = {R^h}^T K R^h,
\end{align}
and the non-affine forces are
\begin{align}
\Xi_{i\mu} &= \frac{\partial^2 E}{\partial \gamma \partial x_{i\mu}} = \frac{\partial}{\partial \gamma}\left(\sum_\alpha \frac{\partial E}{\partial h_\alpha}\frac{\partial h_\alpha}{\partial x_{i\mu}} \right) \\ \nonumber
&= \sum_{\alpha\beta} \frac{\partial h_\alpha}{\partial x_{i\mu}}\frac{\partial^2 E}{\partial h_\alpha \partial h_\beta}\frac{\partial h_\beta}{\partial \gamma} = ({R^h}^T K \dot{\bm{h}})_{i\mu}.
\end{align}
Then the non-affine contribution to the elastic modulus is 
\begin{align}
\bm\Xi^T H^{-1} \bm\Xi = \dot{\bm{h}}^T K R^h ({R^h}^T K R^h)^{-1}{R^h}^T K \dot{\bm{h}}.
\end{align}
Since $K$ is positive definite and diagonal we can scale it out by defining
\begin{align}
\bar{R} = \sqrt{K} R^h
\end{align}
so that
\begin{align}
\bm\Xi^T H^{-1} \bm\Xi = \dot{\bm{h}}^T \sqrt{K} \bar{R} (\bar{R}^T \bar{R})^{-1}\bar{R}^T \sqrt{K} \dot{\bm{h}}.
\end{align}
If the network is not in a critical configuration then $R$ (and hence $\bar{R}$) are full rank, which implies that
\begin{align}
\bar{R} (\bar{R}^T \bar{R})^{-1}\bar{R}^T &= I
\end{align}
so that
\begin{align}
\bm\Xi^T H^{-1} \bm\Xi &= \dot{\bm{h}}^T K \dot{\bm{h}}.
\end{align}
Then if the network does not possess a self-stress, the affine part of the elastic modulus is exactly cancelled by the non-affine part so that all elastic moduli vanish. However, if a network has any self-stresses then $R$ and $\bar{R}$ are not full rank, so those directions get projected out of the non-affine term. Assuming only a single geometric stress $\bm\sigma$ in the left nullspace of $R$ for simplicity, we have
\begin{align}
\bar{R} (\bar{R}^T \bar{R})^{-1}\bar{R}^T &= I - \frac{\bar{\bm\sigma}{\bar{\bm\sigma}}^T}{|\bar{\bm\sigma}|^2},
\end{align}
where $\bar{\bm\sigma} = \sqrt{K}^{-1} \bm\sigma$ is in the left nullspace of $\bar{R}$, since
\begin{align}
\bar{R}^T \bar{\bm\sigma} = \left({R^h}^T \sqrt{K}\right)\left(\sqrt{K}^{-1} \bm\sigma\right) = {R^h}^T \bm\sigma = 0.
\end{align}
Then
\begin{align}
\bm\Xi^T H^{-1} \bm\Xi &= \dot{\bm{h}}^T \sqrt{K} \left(I - \frac{\bar{\sigma}\bar{\sigma}^T}{|\bar{\sigma}|^2}\right) \sqrt{K} \dot{\bm{h}} \\
&= \dot{\bm{h}}^T K \dot{\bm{h}} - \frac{\left(\dot{\bm{h}}^T \bm\sigma \right)^2}{\bm\sigma^T K^{-1} \bm\sigma} \\
&= \dot{\bm{h}}^T K \dot{\bm{h}} - \frac{\left(\sum_\alpha \sigma_\alpha \dot{h}_\alpha \right)^2}{\sum_\alpha \sigma_\alpha^2 /K_{\alpha\alpha}},
\end{align}
so the full elastic modulus is
\begin{align}
\mu_A = \frac{1}{V}\frac{\left(\sum_\alpha \sigma_\alpha \dot{h}_\alpha \right)^2}{\sum_\alpha \sigma_\alpha^2 /K_{\alpha\alpha}} = \frac{1}{V}\frac{\left(\sum_\alpha \sigma_\alpha \dot{h}_\alpha \right)^2}{\sum_\alpha \sigma_\alpha^2 L_\alpha^2/k_{\alpha}}.
\end{align}

\subsection{Numerical Methods for Optimization and Random Sampling on the Critical Manifold}

\subsubsection{Randomly Sampled Configurations}

First, we sample the critical manifold by assigning random rest lengths to the networks and subjecting them to a bulk volumetric strain until they reach the critical manifold and become rigid. We sample the rest lengths from a Gaussian distribution centered at $\ell_\alpha = 1$ with standard deviation $s = 0.3$, which was chosen to give the largest amount of variance in rest lengths while ensuring the networks experience full rigidity percolation at the critical strain.

Once the rest lengths are chosen, we uniformly rescale them as $\ell_\alpha(\gamma) = \ell_\alpha/(1+\gamma)$ (equivalent to a bulk expansion), increasing the strain $\gamma$ from 0 in steps of 0.01. We do this quasistatically, where after each strain step we minimize the energy of the network $E=1/2\sum_\alpha (L_\alpha - \ell_\alpha)^2$. For this we use the FIRE gradient descent algorithm, stopping the minimization once all components of the gradient $\partial E/ \partial x_{i\mu}$ are less than a tolerance of $10^{-16}$. We strain each network until the total energy is at least $10^{-4}$, then record the configuration $x_{i\mu}$ and the geometric stress $\sigma^h_\alpha = (L_\alpha - \ell_\alpha)/L_\alpha$, which we then normalize to $\sum_\alpha (\sigma_\alpha^h)^2 = N_b$ in order to compare with the geometric stresses used in the parameterization.

Rather than randomly assigning rest lengths to the edges, we can uniformly sample in the space of positive-definite geometric stresses and use the geometric stress parameterization to get the corresponding critical configurations. Since geometric stresses are defined only up to an overall scaling, we take each component of the geometric stress from a standard normal distribution: $\sigma^h_\alpha \sim \mathcal{N}(0,1)$, which results in a uniform distribution over the hypersphere of unique geometric stresses. To ensure the resulting critical configurations are rigid, and to compare directly with the strained configurations, we then take the absolute value of the geometric stress ($\sigma^h_\alpha \rightarrow |\sigma^h_\alpha|$), which is a uniform sample over the positive definite orthant of geometric stress space. Once the geometric stress is chosen, we form the prestress matrix $P_{ij} = \sum_\alpha \sigma^h_\alpha g_{\alpha i}g_{\alpha j}$ then remove the first row and column, which eliminates the trivial zero mode by effectively pinning the first vertex at the origin. We then perform an LU decomposition on $P$ and solve for the coordinates of the other vertices using the geometric stress parameterization described in the main text.

\subsubsection{Optimized Configurations}
We also generate four populations of critical configurations by optimizing different objective functions ($V_L$, $V_\tau$, $B_0$, $G_0$) over the space of geometric stresses. To optimize an objective function $O(\bm\sigma^h)$ we use the FIRE gradient descent algorithm, treating the geometric stress as the degrees of freedom, for which we need to calculate the gradient $\partial O / \partial \sigma^h_\alpha$. The normalization of the geometric stress is not relevant for the parameterization, but we constrain the normalization to $\sum_\alpha|\sigma^h_\alpha|^2 = N_b$ so that each $\sigma^h_\alpha$ is $\mathcal{O}(1)$ to prevent any numerical issues. We stop the optimization when the magnitudes of all components of the gradient are less than a threshold value $10^{-12}$.



\end{document}